\documentclass[pdflatex,sn-mathphys]{sn-jnl}% Math and Physical Sciences Numbered Reference Style
\usepackage{tikz}
\usepackage{tikz-3dplot}
\usetikzlibrary{positioning}
\usetikzlibrary{decorations.pathmorphing}
\definecolor{darkgreen}{rgb}{0.0, 0.8, 0.2}

\usepackage{amssymb}
\usepackage{graphicx}%
\usepackage{multirow}%
\usepackage{amsmath,amssymb,amsfonts}%
\usepackage{amsthm}%
\usepackage{mathrsfs}%
\usepackage[title]{appendix}%
\usepackage{xcolor}%
\usepackage{textcomp}%
\usepackage{manyfoot}%
\usepackage{booktabs}%
\usepackage{algorithm}%
\usepackage{algorithmicx}%
\usepackage{natbib}
\usepackage{algpseudocode}%
\usepackage{listings}%
\usetikzlibrary{shapes.geometric}
\usepackage{physics}
\usepackage{subcaption}
\usepackage{diagbox}
\def\commutator#1#2{[#1, #2]}
\usetikzlibrary{decorations.pathreplacing}

\theoremstyle{thmstyleone}%
\theoremstyle{thmstyletwo}%

\theoremstyle{thmstylethree}%

\begin{document}

\title[...]{Key Evaluation of a Relativistic Quantum Key Distribution Over a Turbulent Free-space Optical Channel}


\author*[1]{\fnm{Georgi} \sur{Bebrov}}\email{g.bebrov@tu-varna.bg}

\affil*[1]{\orgdiv{Telecommunications Department}, \orgname{Technical University of Varna}, \orgaddress{\street{1 Studentska Street}, \city{Varna}, \postcode{9010}, \state{Varna}, \country{BULGARIA}}}


\abstract{We investigate the performance of relativistic quantum key distribution (RQKD) over turbulent free-space optical channels for both single-photon and coherent-state implementations. A theoretical model is developed to describe the effect of atmospheric turbulence on interferometric visibility and secret-key generation by incorporating turbulence-induced phase fluctuations into the key-rate analysis. Numerical simulations are performed under realistic channel conditions to evaluate the achievable communication distance and secret-key rate. The results show that atmospheric turbulence degrades the protocol performance by reducing interference visibility, with the severity of the degradation determined by the ratio between the interferometer-arm separation $r$ (or temporal separation of signal and reference states $\Delta T$) and Fried's parameter. The single-photon protocol remains operational over all investigated turbulence regimes, whereas the coherent-state implementation exhibits a critical dependence on the interferometer geometry (especially on $\Delta T$), leading to a threshold beyond which secure key generation is no longer possible. These findings quantify the impact of atmospheric turbulence on relativistic quantum key distribution and provide practical guidelines for the design and optimization of free-space RQKD systems.}

\keywords{quantum key distribution, special relativity}



\maketitle

\section{Introduction}\label{intro}

Quantum key distribution (QKD) enables two legitimate users, conventionally referred to as \textit{Alice} and \textit{Bob}, to establish information-theoretically secure cryptographic keys by exploiting the principles of quantum mechanics. Since the introduction of the BB84 protocol, numerous QKD protocols have been proposed with the aim of improving transmission distance, secret-key generation rate, implementation practicality, and robustness against imperfections. These include decoy-state QKD \cite{hwang,hwang2003,wang2005,wang2}, measurement-device-independent (MDI) QKD \cite{mdi}, twin-field QKD \cite{lucamarini,cal,sns}, phase-matching QKD \cite{pmqkd}, and their numerous variants. The performance of such protocols is commonly assessed through analytical secret-key rate expressions, which provide a quantitative description of the achievable secure key generation under realistic channel conditions and facilitate direct comparison between different protocols.\\
\indent Relativistic quantum key distribution (RQKD) \cite{kravtsov,bebrov} constitutes a fundamentally different class of QKD protocols, whose security relies not only on quantum-mechanical principles but also on relativistic causality. By exploiting the impossibility of superluminal information transfer, RQKD eliminates several assumptions required in conventional QKD security proofs and offers an alternative framework for secure quantum communication. Owing to these distinctive security foundations, RQKD has attracted considerable theoretical interest and has been demonstrated experimentally in proof-of-principle implementations. \\
\indent Despite these advances, the performance analysis of RQKD remains comparatively underdeveloped. Existing studies primarily focus on demonstrating security or experimental feasibility, whereas explicit secret-key rate expressions are generally unavailable. In particular, analytical key-rate models for both the single-photon and weak coherent-state implementations have not been established. This absence hinders a rigorous comparison of RQKD with contemporary QKD protocols and limits the quantitative assessment of its achievable communication distance, channel tolerance, and practical performance under realistic transmission conditions.\\
\indent Furthermore, free-space optical (FSO) \cite{pirandola,fried,dayton} links constitute one of the most promising transmission media for relativistic quantum communication. Their line-of-sight geometry naturally supports relativistic constraints while enabling terrestrial and satellite-based quantum communication. However, atmospheric attenuation, turbulence-induced phase fluctuations, beam wandering, and detector imperfections significantly influence the achievable secret-key generation rate. Consequently, deriving analytical key-rate expressions that explicitly incorporate the characteristics of FSO channels is essential for evaluating the practical performance of relativistic QKD systems.\\
\indent In this work, we introduce analytical secret-key rate expressions for relativistic QKD operating in both the single-photon and weak coherent-state regimes over free-space optical channels. The resulting analytical models enable direct performance evaluation of relativistic QKD under realistic FSO conditions and establish a basis for systematic comparison with existing QKD protocols. The proposed framework provides a quantitative characterization of achievable secret-key rates, transmission distances, and robustness to channel impairments, thereby addressing an important gap in the current literature on relativistic quantum cryptography.\\
\indent The paper is organized as follows. Section \ref{rqkd-protocol} introduces two RQKD protocols: one operating in the single-photon regime (Subsection \ref{protocol}) and another operating in the weak-coherent-state regime (Subsection \ref{coherent-rqkd}). Section \ref{key-rate-evaluation} presents key-rate evaluations of the single-photon RQKD (Subsection \ref{kre-single}) and the coherent-state RQKD (Subsection \ref{kre-coherent}) in the case of a turbulent FSO channel. Section \ref{numerical-results} puts forward numerical results for the key rate of a single-photon RQKD (Subsection \ref{nr-single}) and the key rate of a coherent-state RQKD (Subsection \ref{nr-coherent}). Section \ref{conclusions} outlines the main conclusions of the paper.

\section{Relativistic Quantum Key Distribution}\label{rqkd-protocol}

\subsection{Single-photon Relativistic Quantum Key Distribution}\label{protocol}
Consider a QKD based on the point-to-point setup presented in Fig. \ref{rqkd-fso}. The length $L$ of the link Alice\textendash Bob is preliminary known. The protocol steps of the single-photon RQKD are:
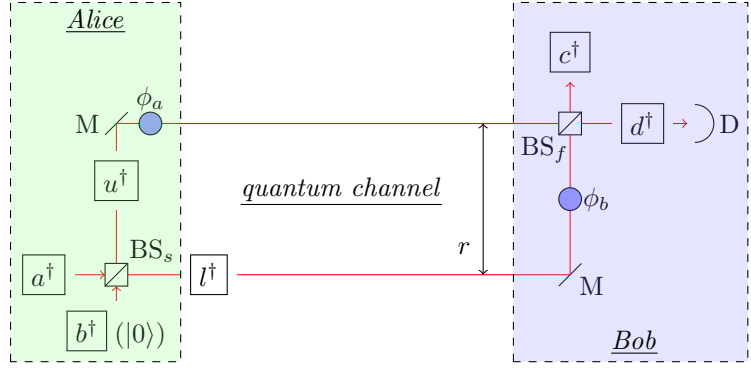
\begin{figure}
	\centering
	\begin{tikzpicture}
		% beam splitters
		\draw (0,0) rectangle (0.3,0.3);
		\draw (0,0) -- (0.3,0.3);
		\node[above right] at (0.2,0.2) {BS$_{s}$};

		\draw (6,2) rectangle (6.3,2.3);
		\draw (6,2) -- (6.3,2.3);
		\node[below left] at (6.25,2.1) {BS$_{f}$}; 
		
		% mirrors
		\draw[black] (0,2) -- (0.3,2.3);
		\draw[black] (6,0) -- (6.3,0.3);
		\node[left] at (0.05,2.15) {M};
		\node[right] at (6.15,0) {M};
		
		% input beam
		\draw[->,red] (-0.4,0.15) -- (0,0.15);
		\draw[->,red] (0.15,-0.2) -- (0.15,0);
		\node[fill=white,left] at (-0.4,0.15) {\fbox{$a^\dagger$}};
		\node[fill=white,below] at (0.15,-0.2) {\fbox{$b^\dagger$}$\;(\ket{0})$};
		% lower arm
		\draw[red] (0.3,0.15) -- (6.15,0.15) -- (6.15,2);
		%\draw[red] (5.7,-1.2) -- (5.7,0);
		\node[fill=white,right] at (1,0.15) {\fbox{$l^\dagger$}};
		
		% upper arm
		\draw[red] (0.15,0.3) -- (0.15,2.15) -- (6,2.15);
		%\draw[red] (0.6,2.3) -- %(5.7,2.3);
		%\draw[red] (5.7,2.3) -- (5.7,0.6);
		\node[fill=white,above] at (0.15,1) {\fbox{$u^\dagger$}};
		
		% phase shifter
		\draw[fill=blue!40] (0.6,2.15) circle (0.15);
		%\node at (3.1,2.9) {$\Delta\phi$};
		\node[above] at (0.6,2.2) {$\phi_a$};
		% phase shifter
		\draw[fill=blue!40] (6.15,1.15) circle (0.15);
		%\node at (3.1,2.9) {$\Delta\phi$};
		\node[right] at (6.2,1.15) {$\phi_b$};
		
		% outputs
		\draw[->,red] (6.15,2.3) -- (6.15,2.7);
		\draw[->,red] (6.3,2.15) -- (7.7,2.15);
		\node[fill=white,right] at (6.7,2.15) {\fbox{$d^\dagger$}};
		\node[fill=white,above] at (6.15,2.7) {\fbox{$c^\dagger$}};

		\draw (7.8,2.4) arc[start angle=90,end angle=-90,radius=0.25];
		%\draw[-] (7.8,1.9) -- (7.8,2.4);
		\node[right] at (8,2.15) {D};

		\draw[dashed,fill=green!50,fill opacity=0.2] (-1.25,-1) rectangle (1,3.75);
		
		\draw[dashed,fill=blue!50,fill opacity=0.2] (5.4,-1) rectangle (8.5,3.75);
		
		%\draw[dashed] (1,-1) -- (1,3.5);
		\node[] at (-0.125,3.5) {\underline{\textit{Alice}}};
		
		%\draw[dashed] (5.25,-1) -- (5.25,3.5);
		\node[] at (7,-0.75) {\underline{\textit{Bob}}};
		
		\node[] at (3.125,1.25) {\underline{\textit{quantum channel}}};
		%\draw[->,red] (6.6,0.45) -- (8.5,0.45);
		%\draw[->,red] (6.3,0.6) -- (6.3,2.2);

		\draw[<->] (5,2.15) -- (5,0.15);
		\node[] at (4.75,0.5) {$r$};
		% output labels
		%\node[right] at (8.5,0.45) {$I_1=\cos^2\frac{\Delta\phi}{2}$};
		%\node[above] at (6.3,2.2) {$I_2=\sin^2\frac{\Delta\phi}{2}$};
	\end{tikzpicture}
	\caption{\label{rqkd-fso} Interferometric setup of single\textendash photon RQKD. BS\textemdash \textit{beamsplitter}; M\textemdash \textit{mirror}; D\textemdash \textit{detector}; $\phi_{a(b)}$\textemdash \textit{phase shift} induced by \textit{Alice} (\textit{Bob}); $r$\textemdash \textit{interferometer arms separation}. Note that a \textit{vacuum} state ($\ket{0}_b$) is fed into input $b$.}
\end{figure}
\begin{figure}[!hbt]
	\centering
	\begin{tikzpicture}[thick]
		
		\draw[color=yellow!20,fill=yellow!20] (0,0) -- (4,4) -- (8,4) -- (4,0) -- cycle;
		\draw[->] (4,0) -- (-2,0) node[above] {$x$};
		\draw[->] (4,0) -- (4,5) node[above] {$ct$};
		\draw[->] (4,0) -- (10,2) node[above] {$y$};
		\node[] at (5.75,1) {$45^\circ$};
		\draw[->] (5,0.3) -| (5,1);
		
		\draw[fill=green!40] (0,0) circle (0.05);
		\draw[fill=green!40] (4,4) circle (0.05);
		\draw[fill=green!40] (8,4) circle (0.05);
		\draw[fill=green!40] (4,0) circle (0.05);
		
		\node[below] at (0,0) {$V$};
		\node[left] at (4,4) {$W$};
		\node[right] at (8,4) {$U$};
		\node[below] at (4,0) {$Q$};

	\end{tikzpicture}
	
	\begin{tikzpicture}[thick]
		
		\draw[color=yellow!20, fill=yellow!20] (0,0) rectangle (6,8);
		
		\draw[dashed,color=orange] (-1,0) -- (6,0);
		\node[left] at (-1,0) {$t_0$};
		
		\draw[dashed,color=orange] (-1,1) -- (6,1);
		\node[left] at (-1,1) {$t_1$};
		
		\draw[dashed,color=orange] (-1,2) -- (6,2);
		\node[left] at (-1,2) {$t_2$};
		
		\draw[dashed,color=orange] (-1,3) -- (6,3);
		\node[left] at (-1,3) {$t_3$};
		
		\draw[dashed,color=orange] (-1,4) -- (6,4);
		\node[left] at (-1,4) {$t_4$};
		
		\draw[dashed,color=orange] (-1,5) -- (6,5);
		\node[left] at (-1,5) {$t_5$};
		
		\draw[dashed,color=orange] (-1,6) -- (6,6);
		\node[left] at (-1,6) {$t_6$};
		
		\draw[dashed,color=orange] (-1,7) -- (6,7);
		\node[left] at (-1,7) {$t_7$};
		
		\draw[dashed,color=orange] (-1,8) -- (6,8);
		\node[left] at (-1,8) {$t_8$};
		
		\draw[fill=green!40] (0,0) circle (0.05);
		\draw[fill=green!40] (0,8) circle (0.05);
		\draw[fill=green!40] (6,8) circle (0.05);
		\draw[fill=green!40] (6,0) circle (0.05);
		
		\node[below] at (0,0) {$V$};
		\node[above] at (0,8) {$W$};
		\node[below right] at (0,8) {$\mathscr{P}$};
		\node[above] at (6,8) {$U$};
		\node[below] at (6,0) {$Q$};
		
		%\draw[color=red] (0,0) -- (6,6);
		
		\draw[color=red] (0.5,-0.5) -- (6,5);
		\draw[color=red] (6,5) -- (5,6);
		
		\draw[color=red] (1,0) -- (0,1);
		\draw[color=red] (0,1) -- (6,7);
		%\draw[color=red] (6,7) -- (5,8);
		
		%M
		\draw (6,4.75) -- (6,5.25);
		\node[right] at (6,5) {M};
		%M
		\draw (0,0.75) -- (0,1.25);
		\node[left] at (0,1) {M};
		
		%BSf
		\node[
		draw,
		diamond,
		minimum width=1,
		minimum height=0.5] at (5,6) {};
		\draw (5,5.75) -- (5,6.25);
		\node[left] at (4.875,6) {BS$_f$};
		
		%PBS
		%\node[
		%draw,
		%diamond,
		%minimum width=1,
		%minimum height=0.5] at (6,7) {};
		%\draw (6,6.75) -- (6,7.25);
		%\node[right] at (6.125,7) {PBS};
		
		%BS
		\node[
		draw,
		diamond,
		minimum width=1,
		minimum height=0.5] at (1,0) {};
		\draw (1,-0.25) -- (1,0.25);
		\node[above] at (1,0.25) {BS$_s$};
		
		%H_a
		\draw[fill=blue!40] (0.5,1.5) circle (0.1);
		\node[above] at (0.45,1.55) {$\phi_a$};
		%H_b
		\draw[fill=blue!40] (5.5,5.5) circle (0.1);
		\node[below] at (5.5,5.5) {$\phi_b$};
		%H_c
		%\draw[fill=blue!40] (5.5,6.5) circle (0.1);
		%\node[above] at (5.25,6.525) {$\hat{H}_c$};
		
		%D
		\draw (5.75,7.125) arc[start angle=135,end angle=-45,radius=0.25];
		%\draw[-] (4.875,7.75) -- (5.25,8.125);
		\node[right] at (6.125,7) {D};
		
		\draw[<->] (3,0) -- (3,1);
		\node[right] at (3,0.5) {$\Delta T$};
		
	\end{tikzpicture}
	
	\caption{\label{spacetime-single} Spacetime diagram of the single-photon RQKD setup. The setup is placed on $\mathscr{P}$ ($VWUQ$) plane. Note that the dashed lines represent planes of constant time $t=t_i$. The time interval $\Delta T$ is a function of the arms separation $r$ ($\Delta T \propto r$). Only the destructive (key-generation, relevant) output of BS$_f$ is depicted.}
\end{figure}
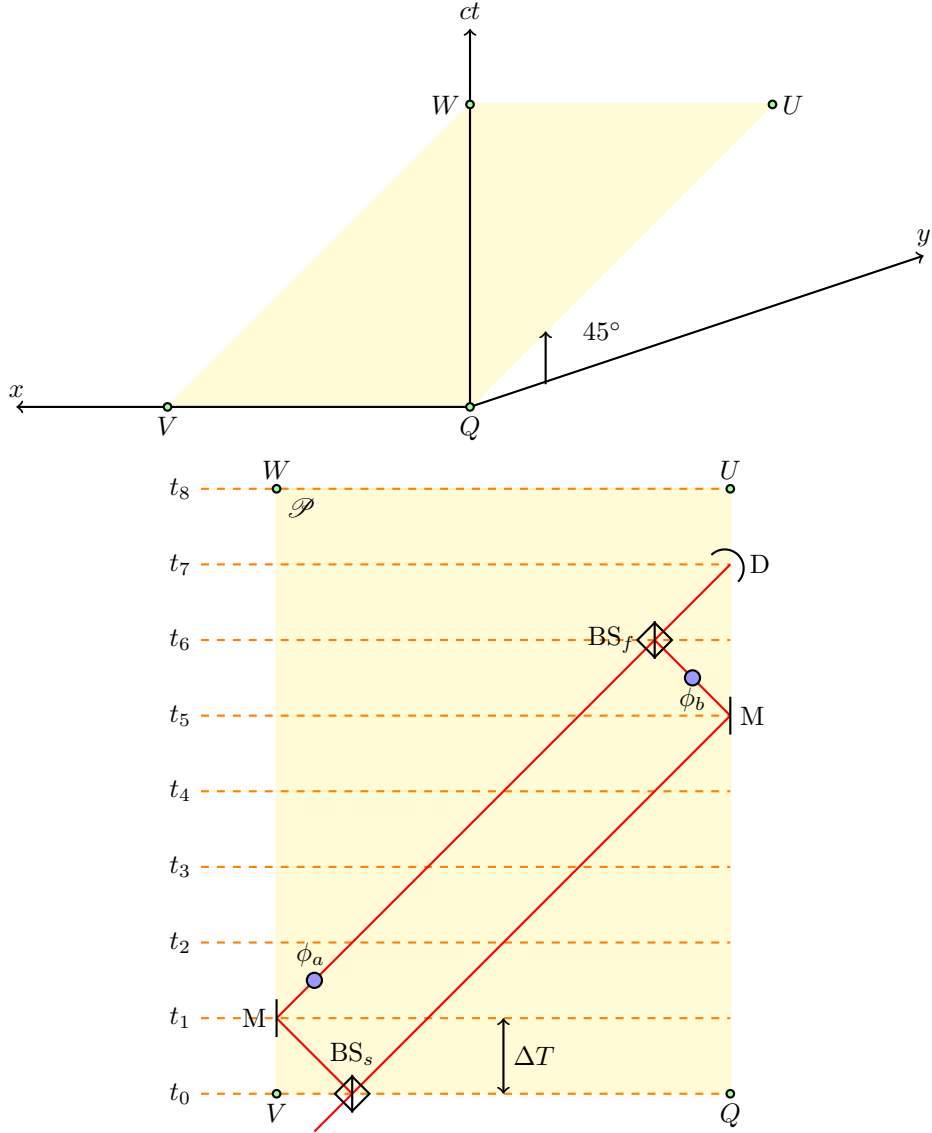
\begin{enumerate}
	\item[\textbf{Step} $0$.] \textit{Alice} prepares $N$ quantum systems $q_k$ ($k=0,\dots,N-1$). We assume $N\rightarrow\infty$, the \textit{asymptotic} case. Each $q_k$ resides in an input mode $a^\dagger$.
	\item[\textbf{Step} $1$.] At time instant $t_k = t_0 + kT$, \textit{Alice} feeds system $q_k$ into the interferometric setup of Fig. \ref{rqkd-fso} (or spacetime diagram of Fig. \ref{spacetime-single}). Note that $t_0$ is considered as the \textit{beginning}\footnote{\textit{protocol beginning}\textemdash it is the time instant at which the first quantum system ($q_0$) is transferred (fed into the setup).} of the protocol (main stage of the protocol) and $T$ is the \textit{transfer period}\footnote{\textit{transfer period}\textemdash it is the time interval between the transfers of two subsequent systems, $q_k$ and $q_{k+1}$.} of the systems. The protocol beginning $t_0$ is preliminary known by \textit{Alice} and \textit{Bob} \cite{kravtsov}. After the beam splitter BS$_s$, each $q_k$ resides in a superposition $(u^\dagger+l^\dagger)\frac{1}{\sqrt{2}}$ (the beam splitter makes the input mode transform as follows: $a^\dagger \rightarrow(u^\dagger+l^\dagger)\frac{1}{\sqrt{2}}$)\footnote{We make use of a beam splitter operator $\hat{B}$ of the form 
	\[\hat{B} = \frac{1}{\sqrt{2}}\begin{bmatrix}
		1 & 1\\ 1 & -1
	\end{bmatrix}.\]
	}. At the upper arm of \textit{her} side, \textit{Alice} applies a \textit{phase shift} operator $\phi_a$ ($\phi_a=\frac{\pi}{2}$ [rad]) according to a probability distribution $\{p_a,1-p_a\}$\textemdash $p_a$ is the probability of applying $\phi_a$. Upon the application of $\phi_a$ the intial upper\textendash arm mode  becomes $\mathrm{e}^{\mathrm{i}\phi_a}u^\dagger$.
	\item[\textbf{Step} $2$.] The quantum systems $\{q_k\}_{k=0}^{N-1}$ (represented by their upper\textendash$\;$and lower\textendash arm modes) are directed towards \textit{Bob} along the interferometric setup. At \textit{his} side, lower arm of the setup, \textit{Bob} applies a \textit{phase shift} operator $\phi_b$ ($\phi_b=\frac{\pi}{2}$ [rad]) according to a probability distribution $\{p_b,1-p_b\}$\textemdash $p_b$ is the probability of applying $\phi_b$. Upon application of $\phi_b$ the intial lower\textendash arm mode becomes $\mathrm{e}^{\mathrm{i}\phi_b}l^\dagger$.
	\item[\textbf{Step} $3$.]  Upper and lower arms of the setup interfere at the second beam splitter BS$_f$. If upper\textendash$\;$and lower\textendash arm modes have identical phases ($\Delta\phi=|\phi_a-\phi_b|=0$), mode $c^\dagger$ appears at the interferometer's output (see Figure \ref{rqkd-fso}). Otherwise ($\Delta\phi\neq 0$), mode $d^\dagger$ appears at the interferometer's output. As shown in Figure \ref{rqkd-fso}, only the events $\Delta\phi\neq 0$ are taken into consideration for key generation\textemdash only mode $d^\dagger$ is registered (mode $d^\dagger$ is directed towards a \textit{detector} D). The detector is usually characterized by efficiency $\eta_{\text{det}}$ and dark count rate $p_{\text{dark}}$. \textit{Bob} discards any  detection event (\textit{i.e.}, a detector click) if it occurs at $t>t_k+\frac{L}{c}$ ($L$\textendash distance between \textit{Alice} and \textit{Bob}, $c$\textendash speed of light)\textemdash such quantum systems (delayed quantum systems) are considered as eavesdropped (intercepted by an eavesdropper). For each non\textendash delayed detector click \textit{Bob} records a key bit (symbol) $B$ according to \textit{his} choice of applying $\phi_b$: $B=0$ if $\phi_b$ is applied; $B=1$ if $\phi_b$ is not applied.
%	\begin{equation*}
%		\begin{split}
%			c^\dagger &= -u^\dagger + l^\dagger \\
%			&= -\frac{1}{\sqrt{2}}\left(\frac{1}{\sqrt{2}}\left(u^\dagger_H+u^\dagger_D\right)-\frac{1}{\sqrt{2}}\left(l^\dagger_H+l^\dagger_D\right)\right) \Rightarrow\\
%			&= -\frac{1}{\sqrt{2}}\left(\frac{1}{\sqrt{2}}\left(u^\dagger_H+u^\dagger_D\right)-\frac{1}{\sqrt{2}}\left(l^\dagger_H+l^\dagger_D\right)\right)
%			\end{split}
%	\end{equation*}
%	Note that the phases related to reflections off mirrors M and OWM are included in the last expression. Next, given the choice of \textit{Bob} to apply $\hat{H}_b$, mode $c^\dagger$ is subjected to an operator $\hat{H}_c$\textemdash $\hat{H}_c$ is applied \textit{if and only if} $\hat{H}_b$ has been applied. 
	\item[\textbf{Step} $4$.] \textit{Bob} announces the time instants (values of $k$) at which \textit{he} obtained non\textendash delayed detections. For each non\textendash delayed detector click (announced by \textit{Bob}) \textit{Alice} records a key bit (symbol) $A$ according to \textit{her} choice of applying $\phi_a$\textemdash $A=1$ if $\phi_a$ is applied; $A=0$ if $\phi_a$ is not applied.\\ 
	\fbox{\color{red}{\textit{Note}:}} In the noiseless (ideal) scenario, a non\textendash delayed detector click occurs when \textit{Alice} and \textit{Bob} are anticorrelated in \textit{their} application of phase shifts.
	\item[\textbf{Step} $6$.] \textit{They} perform \textit{parameter estimation} procedure. \textit{Alice} and \textit{Bob} sacrifice part of \textit{their} key bits in order to determine the quantum bit error rate of the quantum channel. If the error rate does not exceed a preliminary determined threshold, \textit{they} proceed to the next step. Otherwise, \textit{they} terminate the current protocol session and start over the protocol steps.
	\item[\textbf{Step} $6$.] \textit{They} perform \textit{information reconciliation} procedure on \textit{their} keys (key bits).
	\item[\textbf{Step} $7$.] \textit{They} perform \textit{privacy amplification} procedure on \textit{their} reconciled keys (key bits).
\end{enumerate}

\subsection{Coherent-state Relativistic Quantum Key Distribution}\label{coherent-rqkd}

Consider a QKD based on the setup presented in Fig. \ref{coherent-rqkd-fso} (or spacetime diagram of Fig. \ref{spactime-coherent}) \cite{kravtsov}. The length $L$ of the link Alice\textendash Bob is preliminary known. The protocol steps of the coherent-state RQKD are:
\begin{figure}
	\centering
	\begin{tikzpicture}
		% beam splitters
		%\draw (0,0) rectangle (0.3,0.3);
		%\draw (0,0) -- (0.3,0.3);
		%\node[above right] at (0.2,0.2) {BS$_{s}$};

		\draw (6,2) rectangle (6.3,2.3);
		\draw (6,2) -- (6.3,2.3);
		\node[below left] at (6.25,2.1) {BS$_{f}$}; 
		
		% mirrors
		%\draw[black] (0,2) -- (0.3,2.3);
		\draw[black] (6,0) -- (6.3,0.3);
		%\node[left] at (0.05,2.15) {M};
		\node[right] at (6.15,0) {M};
		
		% input beam
		%\draw[->,red] (-0.4,0.15) -- (0,0.15);
		%\draw[->,red] (0.15,-0.2) -- (0.15,0);
		%\node[fill=white,left] at (-0.4,0.15) {\fbox{$a^\dagger$}};
		%\node[fill=white,below] at (0.15,-0.2) {\fbox{$b^\dagger$}$\;(\ket{0})$};
		% lower arm
		\draw[red] (0.15,0.15) -- (6.15,0.15) -- (6.15,2);
		\node[left] at (0.1,0.15) {$\ket{\sqrt{\mu}}_{\textbf{B}'}$};
		%\draw[red] (5.7,-1.2) -- (5.7,0);
		%\node[fill=white,right] at (1,0.15) {\fbox{$l^\dagger$}};
		
		% upper arm
		\draw[red] (0.15,2.15) -- (6,2.15);
		\node[left] at (0.1,2.15) {$\ket{\sqrt{\mu}}_{A'}$};
		%\draw[red] (0.6,2.3) -- %(5.7,2.3);
		%\draw[red] (5.7,2.3) -- (5.7,0.6);
		%\node[fill=white,above] at (0.15,1) {\fbox{$u^\dagger$}};
		
		% phase shifter
		\draw[fill=blue!40] (0.6,2.15) circle (0.15);
		%\node at (3.1,2.9) {$\Delta\phi$};
		\node[above] at (0.6,2.2) {$\varphi_a$};
		% phase shifter
		\draw[fill=blue!40] (6.15,1.15) circle (0.15);
		%\node at (3.1,2.9) {$\Delta\phi$};
		\node[right] at (6.2,1.15) {$\varphi_b$};
		
		% outputs
		\draw[->,red] (6.15,2.3) -- (6.15,2.7);
		\draw[->,red] (6.3,2.15) -- (7.7,2.15);
		%\node[fill=white,right] at (6.7,2.15) {\fbox{$d^\dagger$}};
		%\node[fill=white,above] at (6.15,2.7) {\fbox{$c^\dagger$}};

		\draw (7.65,2.4) arc[start angle=90,end angle=-90,radius=0.25];
		%\draw[-] (7.8,1.9) -- (7.8,2.4);
		\node[right] at (7.8,2.1) {D$_-$};

		\draw (6.4,2.7) arc[start angle=0,end angle=180,radius=0.25];
		%\draw[-] (7.8,1.9) -- (7.8,2.4);
		\node[above] at (6.3,2.9) {D$_+$};

		\draw[dashed,fill=green!50,fill opacity=0.2] (-1.25,-1) rectangle (1,3.75);
		
		\draw[dashed,fill=blue!50,fill opacity=0.2] (5.4,-1) rectangle (8.5,3.75);
		
		%\draw[dashed] (1,-1) -- (1,3.5);
		\node[] at (-0.125,3.5) {\underline{\textit{Alice}}};
		
		%\draw[dashed] (5.25,-1) -- (5.25,3.5);
		\node[] at (7,-0.75) {\underline{\textit{Bob}}};
		
		\node[] at (3.125,1.25) {\underline{\textit{quantum channel}}};
		%\draw[->,red] (6.6,0.45) -- (8.5,0.45);
		%\draw[->,red] (6.3,0.6) -- (6.3,2.2);
		
		% output labels
		%\node[right] at (8.5,0.45) {$I_1=\cos^2\frac{\Delta\phi}{2}$};
		%\node[above] at (6.3,2.2) {$I_2=\sin^2\frac{\Delta\phi}{2}$};
		
		\draw[<->] (5,2.15) -- (5,0.15);
		\node[] at (4.75,0.5) {$r$};
	\end{tikzpicture}
	\caption{\label{coherent-rqkd-fso} Setup of coherent\textendash state RQKD. BS\textemdash \textit{beamsplitter}; M\textemdash \textit{mirror}; D\textemdash \textit{detector}; $\varphi_{a(b)}$\textemdash \textit{phase shift} induced by \textit{Alice} (\textit{Bob}); $r$\textemdash \textit{interferometer arms separation}. System $A'$ identifies the \textit{signal} coherent state, whereas $B'$ identifies the \textit{reference} coherent state. The intensity of the coherent states is denoted by $\mu$. Two detectors are involved in the scheme: detector D$_-$ at which a click takes place if signal and coherent states interfere in a destructive manner (the states are out of phase); detector D$_-$ at which a click takes place if signal and coherent states interfere in a destructive manner (the states are out of phase).}
\end{figure}
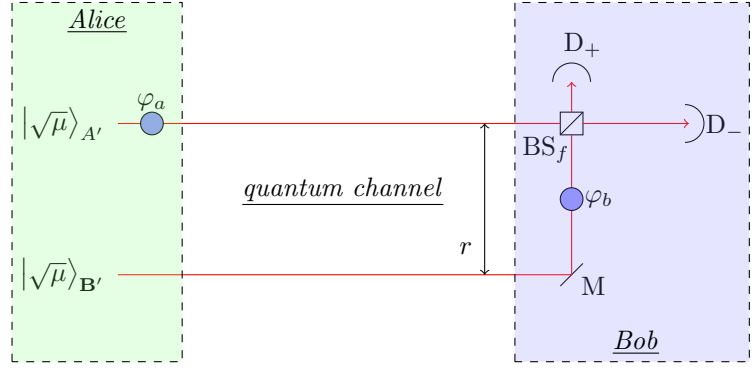
\begin{figure}[!hbt]
	\centering
	\begin{tikzpicture}[thick]
		
		\draw[color=yellow!20,fill=yellow!20] (1.5,0) -- (1.5,5) -- (7,5) -- (7,0) -- cycle;
		\draw[->] (1.5,0) -- (8,0) node[above] {$x$};
		\draw[->] (1.5,0) -- (1.5,6) node[above] {$ct$};
		%\draw[->] (4,0) -- (10,2) node[above] {$y$};
		%\node[] at (5.75,1) {$45^\circ$};
		%\draw[->] (5,0.3) -| (5,1);
		
		\draw[fill=green!40] (1.5,0) circle (0.05);
		\draw[fill=green!40] (1.5,5) circle (0.05);
		\draw[fill=green!40] (7,5) circle (0.05);
		\draw[fill=green!40] (7,0) circle (0.05);
		
		\node[below] at (1.5,0) {$V$};
		\node[left] at (1.5,5) {$W$};
		\node[right] at (7,5) {$U$};
		\node[below] at (7,0) {$Q$};

	\end{tikzpicture}
	
	\begin{tikzpicture}[thick]
		
		\draw[color=yellow!20, fill=yellow!20] (0,0) rectangle (6,8);
		
		\draw[dashed,color=orange] (-1,0) -- (6,0);
		\node[left] at (-1,0) {$t_0$};
		
		\draw[dashed,color=orange] (-1,1) -- (6,1);
		\node[left] at (-1,1) {$t_1$};
		
		\draw[dashed,color=orange] (-1,2) -- (6,2);
		\node[left] at (-1,2) {$t_2$};
		
		\draw[dashed,color=orange] (-1,3) -- (6,3);
		\node[left] at (-1,3) {$t_3$};
		
		\draw[dashed,color=orange] (-1,4) -- (6,4);
		\node[left] at (-1,4) {$t_4$};
		
		\draw[dashed,color=orange] (-1,5) -- (6,5);
		\node[left] at (-1,5) {$t_5$};
		
		\draw[dashed,color=orange] (-1,6) -- (6,6);
		\node[left] at (-1,6) {$t_6$};
		
		\draw[dashed,color=orange] (-1,7) -- (6,7);
		\node[left] at (-1,7) {$t_7$};
		
		\draw[dashed,color=orange] (-1,8) -- (6,8);
		\node[left] at (-1,8) {$t_8$};
		
		\draw[fill=green!40] (0,0) circle (0.05);
		\draw[fill=green!40] (0,8) circle (0.05);
		\draw[fill=green!40] (6,8) circle (0.05);
		\draw[fill=green!40] (6,0) circle (0.05);
		
		\node[below] at (0,0) {$V$};
		\node[above] at (0,8) {$W$};
		\node[below right] at (0,8) {$\mathscr{P}$};
		\node[above] at (6,8) {$U$};
		\node[below] at (6,0) {$Q$};
		
		%\draw[color=red] (0,0) -- (6,6);
		
		\draw[color=red] (1,0) -- (6,5);
		\draw[color=red] (6,5) -- (5,6);
		
		%\draw[color=red] (1,0) -- (0,1);
		\draw[color=red] (0,1) -- (6,7);
		%\draw[color=red] (6,7) -- (5,8);
		
		%M
		\draw (6,4.75) -- (6,5.25);
		\node[right] at (6,5) {M};
		%M
		%\draw (0,0.75) -- (0,1.25);
		%\node[left] at (0,1) {M};
		
		%BSf
		\node[
		draw,
		diamond,
		minimum width=1,
		minimum height=0.5] at (5,6) {};
		\draw (5,5.75) -- (5,6.25);
		\node[left] at (4.875,6) {BS$_f$};
		
		%PBS
		%\node[
		%draw,
		%diamond,
		%minimum width=1,
		%minimum height=0.5] at (6,7) {};
		%\draw (6,6.75) -- (6,7.25);
		%\node[right] at (6.125,7) {PBS};
		
		%BS
		%\node[
		%draw,
		%diamond,
		%minimum width=1,
		%minimum height=0.5] at (1,0) {};
		%\draw (1,-0.25) -- (1,0.25);
		%\node[above] at (1,0.25) {BS$_s$};
		
		%H_a
		\draw[fill=blue!40] (0.5,1.5) circle (0.1);
		\node[above] at (0.45,1.55) {$\varphi_a$};
		%H_b
		\draw[fill=blue!40] (5.5,5.5) circle (0.1);
		\node[below] at (5.5,5.5) {$\varphi_b$};
		%H_c
		%\draw[fill=blue!40] (5.5,6.5) circle (0.1);
		%\node[above] at (5.25,6.525) {$\hat{H}_c$};
		
		%D
		\draw (5.75,7.125) arc[start angle=135,end angle=-45,radius=0.25];
		%\draw[-] (4.875,7.75) -- (5.25,8.125);
		\node[right] at (6.125,7) {D};
		
		\draw[<->] (3,0) -- (3,1);
		\node[right] at (3,0.5) {$\Delta T$};
		
	\end{tikzpicture}
	
	\caption{\label{spactime-coherent} Spacetime diagram of the coherent-state RQKD setup. The setup is placed on $\mathscr{P}$ ($VWUQ$) plane. Note that the dashed lines represent planes of constant time $t=t_i$. The time interval $\Delta T$ is a function of the arms separation $r$ ($\Delta T \propto r$). Only the destructive (key-generation, relevant) output of BS$_f$ is depicted. }
\end{figure}
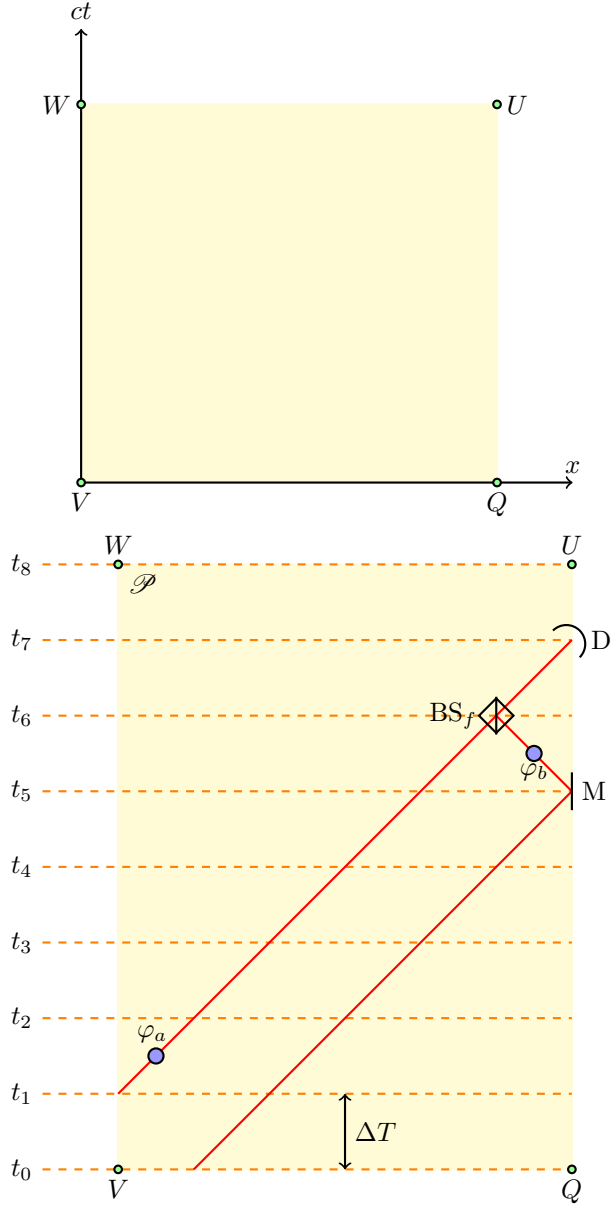
\begin{enumerate}
	\item[\textbf{Step} $0$.] $N$ rounds in total ($N\rightarrow\infty$, \textit{asymptotic} regime) are conducted in the course of a protocol session. For each round, \textit{Alice} chooses the \textit{round} type (\textit{signal} or \textit{test} round) according to \textit{a priori} probability distribution $\{p_{\text{signal}},1-p_{\text{signal}}\}$. 
	\item[\textbf{Step} $1$.] At time instant $t_k = t_0 + kT$ ($k=0,\dots,N$), \textit{Alice} prepares and sends to \textit{Bob} a \textit{reference} coherent state  $\ket{\sqrt{\mu}}_{B'}$ ($\ket{\mathrm{e}^{\mathrm{i}\theta_B}\sqrt{\mu_B}}_{B'}$) in the case of a signal round (in the case of a test round). Note that $t_k$ identifies the beginning of each round, $t_0$ identifies the beginning of the QKD session (first round), and $T$ identifies \textit{transfer period}. The intensity $\mu_B$ and phase $\theta_B$ ($\theta_B\in [0,2\pi)$) are randomly picked, whereas the intensity $\mu$ is preliminary chosen by both \textit{Alice} and \textit{Bob} (\textit{they} both agree on the value of $\mu$ so that the QKD protocol obtains optimal performance).\\
	At time instant $\tau_k = t_k + \Delta T$ ($\Delta T$ denotes the temporal separation between the reference and signal states), \textit{Alice} prepares and sends to \textit{Bob} a \textit{signal} coherent state $\ket{\mathrm{e}^{\mathrm{i}\varphi_A}\sqrt{\mu}}_{A'}$ ($\ket{\mathrm{e}^{\mathrm{i}\theta_A}\sqrt{\mu_A}}_{A'}$) in the case of a signal round (in the case of a test round). The intensity $\mu_A$ and phase $\theta_A$ ($\theta_A\in [0,2\pi)$) are randomly picked. The signal phase $\varphi_A$ ($\varphi_A\in\{0,\pi\}$ [rad]) is picked in a random manner.
	\item[\textbf{Step} $2$.] At \textit{his} side, lower arm of the setup, \textit{Bob} randomly applies a \textit{phase shift} $\varphi_b$ ($\varphi_b\in\{0,\pi\}$ [rad]) to the reference coherent state\textemdash the reference state thus takes the form $\ket{\mathrm{e}^{\mathrm{i}\varphi_B}\sqrt{\mu}}_{B'}$ ($\ket{\mathrm{e}^{\mathrm{i}(\theta_B+\varphi_B)}\sqrt{\mu_B}}_{B'}$) in the case of a signal round (in the case of a test round). As shown in Figure 
	\ref{coherent-rqkd-fso}, the reference state undergoes a reflection from \textit{mirror} M. Therefore the reference state becomes $\ket{\mathrm{e}^{\mathrm{i}\varphi_B}\mathrm{e}^{\mathrm{i}\pi}\sqrt{\mu}}_{B'} = \ket{\mathrm{e}^{\mathrm{i}(\varphi_B+\pi)}\sqrt{\mu}}_{B'}$ ($\ket{\mathrm{e}^{\mathrm{i}(\theta_B+\varphi_B+\pi)}\sqrt{\mu_B}}_{B'}$). \\
	\fbox{\color{red}{\textit{Note}:}} The reflection from the mirror gives rise to a non-zero phase difference ($|\varphi_a - \varphi_b|\neq 0$) despite \textit{Alice}'s and \textit{Bob}'s correlated phase choices.
	\item[\textbf{Step} $3$.]  Signal and reference coherent states interfere at the beam splitter BS. In the signal rounds, a destructive interference ($\Delta\varphi=|\varphi_a-\varphi_b|=\pi$, $\Delta\varphi$ is the phase difference) between signal and reference states is of interest. The destructive interference leads to a possible registration of a click at the \textit{detector} D\textemdash in other words, it is used for a key generation. In the test rounds, detectors at both constructive and destructive outputs of the BS are used. The detector is usually characterized by efficiency $\eta_{\text{det}}$ and dark count rate $p_{\text{dark}}$. \textit{Bob} discards any  detection event (\textit{i.e.}, a detector click) if it occurs at $t>t_k+\frac{L}{c}$ ($L$\textendash distance between \textit{Alice} and \textit{Bob}, $c$\textendash speed of light)\textemdash such quantum systems (delayed quantum systems) are considered as eavesdropped (intercepted by an eavesdropper).
	\item[\textbf{Step} $4$.] \textit{Bob} announces the time instants (values of $k$) at which \textit{he} obtained non\textendash delayed detections. For each non\textendash delayed detector click (announced by \textit{Bob}) in a signal round, \textit{Alice} records a key bit (symbol) $A$ according to \textit{her} choice of applying $\phi_a$\textemdash $A=1$ if $\varphi_a=\pi$ is applied; $A=0$ if $\varphi_a=0$.\\ 
	\fbox{\color{red}{\textit{Note}:}} In the noiseless (ideal) scenario, a non\textendash delayed detector click occurs when \textit{Alice} and \textit{Bob} are correlated in \textit{their} phase-shift choices.\\
	\textit{Alice} announces the type of each round. For each non\textendash delayed detector click in a signal round, \textit{Bob} records a key bit (symbol) $B$ according to \textit{his} choice of $\varphi_b$: $B=0$ if $\varphi_b=0$; $B=1$ if $\varphi_b=\pi$.
	\item[\textbf{Step} $6$.] \textit{They} perform \textit{parameter estimation} procedure. \textit{Alice} and \textit{Bob} sacrifice part of \textit{their} key bits in order to determine the quantum bit error rate of the quantum channel. If the error rate does not exceed a preliminary determined threshold, \textit{they} proceed to the next step. Otherwise, \textit{they} terminate the current protocol session and start over the protocol steps.
	\item[\textbf{Step} $6$.] \textit{They} perform \textit{information reconciliation} procedure on \textit{their} keys (key bits).
	\item[\textbf{Step} $7$.] \textit{They} perform \textit{privacy amplification} procedure on \textit{their} reconciled keys (key bits).
\end{enumerate}

\section{Key-rate evaluation}\label{key-rate-evaluation}

\subsection{Single-photon regime}\label{kre-single}
In the \textit{lossless} scenario (noisy, but lossless quantum channel), the key rate of the single-photon RQKD protocol is defined by the general expression \cite{bebrov}
\begin{equation}
	\begin{split}
		R &=  p_{\text{det}}\left(1 - \chi - fH(e)\right).
	\end{split}
\end{equation}
In the \textit{lossless} scenario\textemdash the quantum channel transmissivity is $\eta_{\text{t}}=1$ and detector efficiency is $\eta_{\text{det}}=1$. Here $p_{\text{det}}$ denotes the probability of \textit{Bob} to register a detection click (key\textendash generation event) in the proposed protocol (see Appendix \ref{math-single-photon}), $e$ is the \textit{quantum bit error rate}, $f$ is the \textit{error correction efficiency}, $H(\bullet)$ is the Shannon entropy, and $\chi$ is the information leaked to \textit{Eve}, an eavesdropper. The single\textendash photon error rate $e$ is given by \cite{ma2005}
\begin{equation}
	\begin{split}
		e = \frac{e_0p_{\text{dark}}+e_{\text{det}}\left(\eta\right)}{Y_1} = \frac{e_0p_{\text{dark}}+e_{\text{det}}\left(\eta_{\text{t}}\eta_{\text{det}}\right)}{Y_1},
	\end{split}
\end{equation}
where $e_{\text{det}}$ is the probability that a photon hit the erroneous detector (in this work, this is the probability that a photon is erroneously directed by final BS, $e_{\text{det}}=0.03+\langle p'_{\text{error}}\rangle$, where $0.03$ is contributed by setup/devices imperfections \cite{ma2005} and $\langle p'_{\text{error}}\rangle$ is contributed by channel turbulence (see Appendix \ref{math-single-photon})), $e_0$ is the background error rate ($e=\frac{1}{2}$\textemdash consider a random background), $p_{\text{dark}}$ is the detector dark count, and $Y_1$ is the single\textendash photon yield. Note that the \textit{overall} transmissivity $\eta$ is given by the product $\eta_{\text{t}}\eta_{\text{det}}$. The yield is given by \cite{ma2005}
\begin{equation}
	\begin{split}
		Y_1 = p_{\text{dark}} + \eta - p_{\text{dark}}\eta.
	\end{split}
\end{equation}
In case of a noisy and lossy quantum channel, the key rate takes the form
\begin{equation}\label{rate-loss}
	\begin{split}
		R' =  Y_1p_{\text{det}}\left(1 - \chi - fH(e)\right).
	\end{split}
\end{equation}
%\indent In QKD protocols, the Holevo information $\chi$ is used to represent the information leakage to \textit{Eve}, an eavesdropper.
As turbulent FSO channel is considered, the overall transmissivity takes the form $\eta=\eta_{\text{lt}}\eta_{\text{det}}\eta_{\text{atm}}$ (see for reference Tab. \ref{table-parameters}), where $\eta_{\text{lt}}\eta_{\text{atm}}$ denotes the transmissivity of the turbulent channel. \\
\indent As presented in the protocol steps (Sec. \ref{protocol}) and shown in Appendix \ref{math-single-photon}, \textit{Alice} encodes \textit{her} bits in relative phases $0$ and $\frac{\pi}{2}$. Here we consider the worst\textendash case scenario\textemdash we assume that \textit{Eve} has an access to the upper bound of $\chi$, known as \textit{Holevo information} or \textit{Holevo bound}. In this scenario, an ensemble $\{\frac{1}{2}, \ket{\psi_0}; \frac{1}{2}, \psi_{\frac{\pi}{2}}\}$ completely characterizes the information source of \textit{Alice} (\textit{Alice}'s key bits), where 
\begin{equation*}
	\begin{split}
		\ket{\psi_0} &= \frac{1}{\sqrt{2}}\left(u^\dagger +l^\dagger\right)\ket{0},\\
		\ket{\psi_\frac{\pi}{2}} &= \frac{1}{\sqrt{2}}\left(\mathrm{e}^{\mathrm{i}\frac{\pi}{2}}u^\dagger +l^\dagger\right)\ket{0}.
	\end{split}
\end{equation*}
%\textit{Eve} is assumed perform a state discrimination on an ensemble $\{\frac{1}{2}, H; \frac{1}{2}, D\}$ in order to learn the key bits of \textit{Alice}.  
Take into account that we consider \textit{Eve} having only access to the quantum channel (states transferred from \textit{Alice} to \textit{Bob}). This ensemble is related to the following density operator   
\begin{equation}\label{sigma}
	\begin{split}
		\sigma = \frac{1}{2}\ketbra{\psi_0} + \frac{1}{2}\ketbra{\psi_\frac{\pi}{2}} = p_0\sigma_0 + p_1\sigma_1.
	\end{split}
\end{equation} 
The Holevo information $\chi$ for the ensemble of concern has the form \cite{lima,lin-lutken}
\begin{equation*}
	\begin{split}
		\chi &= S(\sigma) - \sum\limits_{i=0}^{1} p_iS(\sigma_i) \\
		& = S(\sigma) = H(\{\lambda_i\}) = 0.601,
	\end{split}
\end{equation*}
where $S(\bullet)$ is the von Neumann entropy, $\{\lambda_i\}$ are the eigenvalues of $\sigma$, and $H(\{\lambda_i\})$ is given by
\begin{equation*}
	\begin{split}
		H(\{\lambda_i\}) = -\lambda_0\log_2(\lambda_0) -  \lambda_1\log_2(\lambda_1).
	\end{split}
\end{equation*}
Note that
\begin{equation*}
	\begin{split}
		\lambda_0&=\frac{1}{2}\left(1+\frac{1}{\sqrt{2}}\right),\\ \lambda_1&=\frac{1}{2}\left(1-\frac{1}{\sqrt{2}}\right).
	\end{split}
\end{equation*}
Notice that $S(\sigma_i)=0,\;\forall i$ since $\sigma_i$ (see Eq. \eqref{sigma}) are pure states (see Appendix \ref{pure}). Take into account that
\begin{equation}
	\begin{split}
		\braket{\psi_0}{\psi_{\frac{\pi}{2}}} &= \frac{1+\mathrm{i}}{2}.
	\end{split}
\end{equation}
We make use of relations \eqref{commutators} in the last expression. Hence
\begin{equation}
	\begin{split}
		\abs{\braket{\psi_0}{\psi_{\frac{\pi}{2}}}} = \frac{1}{\sqrt{2}}.
	\end{split}
\end{equation}

\subsection{Weak-coherent-state regime}\label{kre-coherent}
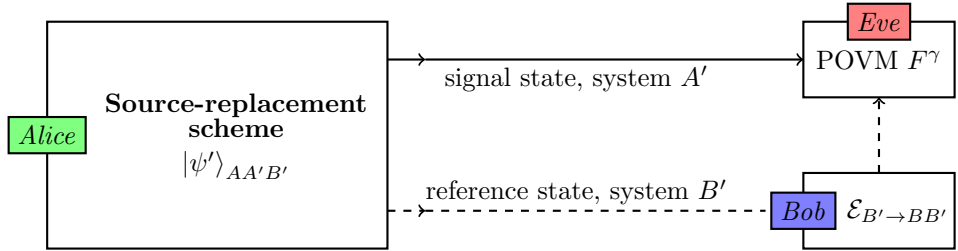
\begin{figure}[!hbt]
	\centering
	\begin{tikzpicture}[thick]
		\draw[draw=black] (-6.5,-1.5) rectangle (-2,1.5);
		\node[draw,rectangle,fill=green!50] at (-6.5,0) {\textit{Alice}};
		\draw[->] (-2,1) -- (-1.5,1);
		\draw[->,dashed] (-2,-1) -- (-1.5,-1);
		\draw[->] (-1.5,1) -- (3.5,1);
		\draw[-,dashed] (-1.5,-1) -- (3,-1);
		\draw[draw=black] (3.5,-1.5) rectangle (5.5,-0.5);
		\node[draw,rectangle,fill=blue!50] at (3.5,-1) {\textit{Bob}};
		\node[] at (4.75,-1) {$\mathcal{E}_{B'\rightarrow BB'}$};
		\draw[->,dashed] (4.5,-0.5) -- (4.5,0.5);
		\draw[draw=black] (3.5,0.5) rectangle (5.5,1.5);
		\node[draw,rectangle,fill=red!50] at (4.5,1.5) {\textit{Eve}};
		\node[] at (4.5,1) {POVM $F^\gamma$};

		\node[] at (-4,0.4) {\textbf{Source-replacement }};
		\node[] at (-4,0.1) {\textbf{scheme}};
		\node[] at (-4,-0.4) {$\ket{\psi'}_{AA'B'}$};
		\node[] at (0.5,0.75) {signal state, system $A'$};
		\node[] at (0.5,-0.75) {reference state, system $B'$};
		
	\end{tikzpicture}
	\caption{\label{weak-coherent-model} Security analysis model in the case of weak-coherent-state regime. \textit{Alice} prepares signal $A'$ and reference $B'$ coherent states. The signal state is sent directly to \textit{Eve}, whereas the reference state is first sent to \textit{Bob} for modulation and subsequently forwarded to \textit{Eve}. \textit{Eve} has complete control over the detection system\textemdash the worst scenario is considered. \textit{Eve} performs POVM $F^\gamma$ (joint measurement) on both signal and modulated reference states (system $A'B'$).}
\end{figure}

The key rate evaluation (security analysis) in this regime is based on the framework of Ref. \cite{lin-lutken} (Lin\textendash L\"{u}tkenhaus framework). The analysis follows the model of Figure \ref{weak-coherent-model}. It consists of applying the \textit{Source-replacement scheme} \cite{ferenczi,lin-lutken} followed by invoking the \textit{Devetak-Winter formula} \cite{dev}. \\
\indent  Based on the \textit{source-replacement} approach, \textit{Alice} prepares the following state
\begin{equation}
	\begin{split}
		\ket{\psi}_{AA'BB'} = \sum\limits_{x,y} \sqrt{p_x}\sqrt{p_y}\ket{x}_A\ket{\varphi_x}_{A'}\otimes \ket{y}_B\ket{\varphi_y}_{B'}.
	\end{split}
\end{equation} 
According to \textit{a priori} probability distribution $\{p_x\}$ \textit{Alice} chooses a state from the set of possible signal states $\{\ket{\varphi_x}\}$ for system $A'$. The classical register $A$ records the signal choice of \textit{Alice}. Above we assume that a system $B'$ (initially set in state $\ket{+\sqrt{\mu}}$) travels to \textit{Bob} and \textit{he} modulates (encodes) it into
\begin{equation*}
	\begin{split}
		\sum\limits_{y}\sqrt{p_y}\ket{y}_B\ket{\varphi_y}_{B'}
	\end{split}
\end{equation*}
where a signal state $\ket{\varphi_y}$ (an element of the set $\{\ket{\varphi_y}\}$) is prepared according to \textit{a priori} probability distribution $\{p_y\}$. The register $B$ records the signal choice $y$ of \textit{Bob}. \\
\indent The above lines can be represented as follows. \textit{Alice} prepares
\begin{equation}
	\begin{split}
		\ket{\psi'}_{AA'B'} = \sum\limits_{x} \sqrt{p_x}\ket{x}_A\ket{\varphi_x}_{A'}\otimes \ket{\varphi_y=\varphi_0}_{B'}.
	\end{split}
\end{equation} 
Initially system $B'$ is set in a default state: $\ket{\varphi_y}=\ket{\varphi_0}=\ket{\mathrm{e}^{\mathrm{i}0}\sqrt{\mu}}=\ket{+\sqrt{\mu}}$. Note that system $B'$ represents the \textit{reference state} sent to \textit{Bob} by \textit{Alice} in a RQKD protocol.\\
\indent Next \textit{Alice} sends systems $A'$ and $B'$ to \textit{Eve} and \textit{Bob}, respectively. According to the RQKD protocol, \textit{Bob} performs phase encoding on system $B'$. In other words \textit{he} applies a completely-positive and trace-preserving ($\mathsf{cptp}$) map to system $B'$. The encoding $\mathsf{cptp}$ map of \textit{Bob} has the form
\begin{equation}\label{cptp}
	\begin{split}
		\mathcal{E}_{AA'B'\rightarrow AA'BB'}(\ketbra{\psi'}_{AA'B'}) = \mathcal{I}_{AA'}\otimes \mathcal{E}_{B'\rightarrow BB'}(\ketbra{\psi'}_{AA'B'})
	\end{split}
\end{equation}
$\mathcal{I}$ being the \textit{identity} channel and 
\begin{equation}
	\begin{split}
		\mathcal{E}_{B'\rightarrow BB'}(\rho) = \sum\limits_k K_k \rho K^\dagger_k 
	\end{split}
\end{equation} 
where
\begin{equation}
	\begin{split}
		K_k &= (\sqrt{p_0}\ket{0}_B\ket{\varphi_0=-\sqrt{\mu}}_{B'} + \sqrt{p_1}\ket{1}\ket{\varphi_1=+\sqrt{\mu}}_{B'})\bra{e_k}_{B'}\\
		&=\ket{\zeta}_{BB'}\bra{e_k}_{B'}
	\end{split}
\end{equation}
where $\{\ket{e_k}\}$ is an orthonormal basis of $B'$ $\left(\sum_k \ketbra{e_k} = I \right)$. It is straightforward that the $\mathsf{cptp}$ map of $\eqref{cptp}$ leads to
\begin{equation*}
	\begin{split}
		\ket{\psi}_{AA'BB'} = \sum\limits_{x,y} \sqrt{p_x}\sqrt{p_y}\ket{x}_A\ket{\varphi_x}_{A'}\otimes \ket{y}_B\ket{\varphi_y}_{B'}.
	\end{split}
\end{equation*} 
Take into account that a phase shift of $\pi$ [rad] is induced into the coherent state of system $B'$ in the above $\mathsf{cptp}$ map\footnote{In principle, $\varphi_0$ is equal to $\sqrt{\mu}$, whereas $\varphi_1$ is equal to $-\sqrt{\mu}$. However, in the $\mathsf{cptp}$ introduced above, $\varphi_0=-\sqrt{\mu}=\mathrm{e}^{\mathrm{i}\pi}\sqrt{\mu}$ and $\varphi_1=\sqrt{\mu}=-\mathrm{e}^{\mathrm{i}\pi}\sqrt{\mu}$, where $\mathrm{e}^{\mathrm{i}\pi}$ is the induced phase shift due to reflection off a mirror}. The phase shift is due to reflection off the mirror present in the setup of Figure \ref{coherent-rqkd-fso}.\\
\indent According to the Lin\textendash L\"{u}tkenhaus framework \cite{lin-lutken}, $\ket{\psi}_{AA'BB'}$ is now subjected to a POVM attack by \textit{Eve}. The attack is defined by the $\mathsf{cptp}$ map \cite{lin-lutken}
\begin{equation}
	\begin{split}
		\mathcal{E}_{A'B'\rightarrow EC}(\rho) = \sum\limits_{\gamma}p_{\gamma}\left(\sqrt{F^\gamma}\rho\sqrt{F^\gamma}\right)_E\otimes \ketbra{\gamma}_C
	\end{split}
\end{equation}
$\gamma$ being the measurement outcome announced by \textit{Eve}, the classical register $C$ is used to record $\gamma$, and $F^\gamma$ are the POVM operators used to perform a measurement on system $A'B'$. Applied to system $AA'BB'$, the last $\mathsf{cptp}$ map leads to
\begin{equation}\label{state-attack}
	\begin{split}
		\rho_{ABEC} &= \mathcal{I}_{AB}\otimes\mathcal{E}_{A'B'\rightarrow EC}\left(\ketbra{\psi}_{AA'BB'}\right)\\
		&=\sum\limits_{x,y,x',y'}\sqrt{p_xp_{x'}p_yp_{y'}}\ket{x,y}\bra{x',y'}_{AB}\\
		&\otimes\sum\limits_{\gamma}p_{\gamma}\left(\sqrt{F^\gamma}\ket{\varphi_x,\varphi_y}\bra{\varphi_{x'},\varphi_{y'}}\sqrt{F^\gamma}\right)_E\otimes \ketbra{\gamma}_C.
	\end{split}
\end{equation}
Recall that only $\gamma=-$ outcome is used for key generation (see protocol steps of RQKD in the weak-coherent-state regime). Thus a measurement outcome sifting takes place. The sifting represents applying a projection operator of the form
\begin{equation}
	\begin{split}
		\Pi_{\text{outcome}} = \mathcal{I}_{ABE}\otimes \ketbra{\gamma=-}_C.
	\end{split}
\end{equation}
Therefore 
\begin{equation}
	\begin{split}
		&\Pi_{\text{outcome}}\rho_{ABEC}\Pi_{\text{outcome}} = \rho'_{ABEC} \\
		&=\sum\limits_{x,y,x',y'}\sqrt{p_xp_{x'}p_yp_{y'}}\ket{x,y}\bra{x',y'}_{AB}\otimes\left(\sqrt{F^-}\ket{\varphi_x,\varphi_y}\bra{\varphi_{x'},\varphi_{y'}}\sqrt{F^-}\right)_E\\
		&\otimes\ketbra{\gamma=-}_C 
	\end{split}
\end{equation}
Note that $x=y$ leads to $\gamma=-$ (as noiseless case is considered).\\
\indent We now proceed to evaluate the secret key rate. For this purpose we first apply a \textit{key} map to $\rho'_{ABEC}$ and then adopt Devetak-Winter formula.\\
\indent To establish key bits, \textit{Alice} and \textit{Bob} perform POVMs $M_A$ and $M_B$, respectively\textemdash \textit{Alice} performs measurement $M_A$ ($M_A=\{\ketbra{x}_A\}$) on system $A$, whereas \textit{Bob} performs measurement $M_B$ ($M_B=\{\ketbra{y}_B\}$) on system $B$. The measurement outcome of $M_A$ ($M_B$) is stored in a register $X$ ($Y$). Then a key map $\mathcal{K}$ is applied to system $X$. The map is trivial and characterized by the transform
\begin{equation}
	\begin{split}
		\mathcal{K}_{X\rightarrow K}(\rho_X) = I_X(\ketbra{x}_X)I_X = \ketbra{k}_K.
	\end{split}
\end{equation}
The overall result of performing POVMs ($M_A$ and $M_B$) and applying $\mathcal{K}$ is
\begin{equation}
	\begin{split}
		\rho_{KYEC} = \sum\limits_{k,y}p_{\gamma=-}\times p_{k,y|\gamma=-}\ketbra{k}_K\otimes\ketbra{y}_Y\otimes\rho_E^{k,y,\gamma=-}\otimes\ketbra{\gamma=-}_C
	\end{split}
\end{equation}
where $\rho_E^{k,y,\gamma=-}$ is the state of \textit{Eve} conditioned on $k,y,\gamma=-$. Note that $p_{k,y,\gamma=-}=p_{\gamma=-}p\times _{k,y|\gamma=-}$. \\
\indent The Devetak-Winter formula is used to determine the secret key rate of a QKD protocol under collective attacks and asymptotic regime of operation. The Devetak-Winter formula is applied to state $\rho_{KYEC}$ shared between \textit{Alice}, \textit{Bob}, and \textit{Eve}. Take into account that $\rho_{KYEC}$ can be recast as
\begin{equation}
	\begin{split}
		\rho_{KYEC} = p_{\gamma=-}\times\rho_{KYE}^{\gamma=-}\otimes\ketbra{\gamma=-}_C
	\end{split}
\end{equation}
where the conditional shared by \textit{Alice}, \textit{Bob}, and \textit{Eve}
\begin{equation}
	\begin{split}
		\rho_{KYE}^{\gamma=-}=\sum\limits_{k,y}p_{k,y|\gamma=-}\ketbra{k}_K\otimes\ketbra{y}_Y\otimes\rho_E^{k,y,\gamma=-}
	\end{split}
\end{equation}
is conditioned on $\gamma$. According to Devetak-Winter formula, given announcement $\gamma$, the number of secret key bits that can be extracted out of $\rho_{KYE}^{\gamma=-}$ is
\begin{equation}
	\begin{split}
		r(\rho_{KYE}^{-}) = I(K:Y)_{\rho_{KYE}^{-}} - I(K:E)_{\rho_{KYE}^{-}} = 1 - \delta^-_{\text{EC}} - \chi(K:E)_{\rho_{KYE}^{-}}
	\end{split}
\end{equation}
where $\delta^-_{\text{EC}}$ is the average amount of information leakage during the error correction procedure of a QKD given $\gamma=-$. This quantity is calculated by
\begin{equation}\label{ec}
	\begin{split}
		\delta^-_{\text{EC}} = f_{\text{EC}}\times H(e^-)
	\end{split}
\end{equation}
$f_{\text{EC}}$ being the efficiency of the error correction, $e^-$ being the error rate in case of $\gamma=-$, and $H(\bullet)$ being the Shannon entropy.\\
\indent The quantity $\chi(K:E)_{\rho_{KYE}^{-}}$ is the \textit{Holevo information} which is defined by \cite{lin-lutken}
\begin{equation}\label{holevo}
	\begin{split}
		\chi(K:E)_{\rho_{KYE}^{\gamma}} = S(\rho_E^\gamma) - \sum\limits_k p_{k|\gamma}\times S(\rho_E^{k,\gamma})
	\end{split}
\end{equation}
where $S(\bullet)$ denotes the von Neumann entropy. Note that 
\begin{equation}\label{cond-states}
	\begin{split}
		\rho_E^{k,\gamma=-}&=\sum\limits_y p_{y|k,\gamma=-}\times\rho_E^{k,y,\gamma=-},\\
		\rho_E^{\gamma=-} &= \sum\limits_k p_{k|\gamma=-}\times\rho_E^{k,\gamma=-}.
	\end{split}
\end{equation}
The state $\rho_E^{k,y,\gamma=-}$ is expressed as
\begin{equation}
	\begin{split}
		\rho_E^{k,y,\gamma=-} = \ketbra{\Theta_{k,y}^{\gamma=-}}
	\end{split}
\end{equation}
following equation \eqref{state-attack}, as shown in Ref.\cite{lin-lutken}. The conditional state of \textit{Eve} $\ket{\Theta_{k,y}^{\gamma=-}}$ is given by
\begin{equation}
	\begin{split}
		\ket{\Theta_{k,y}^{\gamma=-}} = \frac{\sqrt{F^-}\ket{\varphi_k,\varphi_y}}{\sqrt{\bra{\varphi_k,\varphi_y}F^-\ket{\varphi_k,\varphi_y}}}.
	\end{split}
\end{equation}
The total amount of secret key bits that \textit{Alice} and \textit{Bob} can extract out of $\rho_{KYEC}$ is given by \cite{lin-lutken}
\begin{equation}\label{overall-coherent-rate}
	\begin{split}
		R(\rho_{KYEC}) = \sum\limits_\gamma p_\gamma \times r(\rho_{KYE}^{\gamma}) = p_{\gamma=-} \times r(\rho_{KYE}^{\gamma=-}).
	\end{split}
\end{equation}

\section{Numerical results}\label{numerical-results}
\subsection{Numerical results of single-photon RQKD}\label{nr-single}
\begin{table}[!hbt]
	\caption{\label{table-parameters} Parameters used in evaluating the key rate of a single-photon (and coherent-state) RQKD over a turbulent FSO. Values and expressions below are taken in accordance with Refs. \cite{fried,dayton,anguita,peters,cardoso,parra,liao}. }
	\begin{tabular}{c|c}
		\hline
		\textbf{Quantity} & \textbf{Value}/\textbf{Expression}\\
		\hline\hline
		\textit{Wavenumber} & $k=\frac{2\pi}{\lambda}$; $\lambda$ measured in [m]\\
		\textit{Wavelength} & $\lambda=850\times 10^{-9}$[m] ($850$[nm])\\
		\textit{Rytov number} & $\sigma_{R}^2 = 1.23C^2_nk^{\frac{7}{6}}L^{\frac{11}{6}}$;  $L$ measured in [m]\\
		\textit{Stucture constant} & $C^2_n = 10^{-18}$ \\
		\textit{Transmissivity} & $\eta=\eta_{\text{det}}\eta_{\text{lt}}\eta_{\text{atm}}$\\
		\textit{Detector efficiency} & $\eta_{\text{det}}=0.5$\\
		\textit{Dark count rate} & $p_{\text{dark}} = 10^{-10}$\\
		\textit{Turbulent link transmissivity} & $\eta_{\text{lt}} = 1-\mathrm{e}^{-
			\frac{2a_R^{2}}{w_{\text{lt}}^2}}$; $a_R$ measured in [m]\\
		\textit{Atmospheric loss} & $\eta_{\text{atm}}=\mathrm{e}^{-L\alpha_0\mathrm{e}^{-\frac{h_0}{6600}}}$; $h_0$, $L$ measured in [m]\\
		\textit{Receiver aperture radius} & $a_R=0.05$[m] ($5$[cm])\\
		\textit{Long-term beam waist at the receiver} & $w_{\text{lt}}=w_L\sqrt{1+\frac{4}{3}q\times\Lambda}$\\
		\textit{Beam's spot size} & $w_L=\sqrt{w_0^2\left[\left(1-\frac{L}{R_0}\right)^2+\left(\frac{L}{L_R}\right)^2\right]}$; $L$ measured in [m]\\
		\textit{Beam's Rayleigh length} & $L_R =\frac{\pi w_0^2}{\lambda}$; $w_0$ measured in [m]\\
		\textit{Initial spot size} (\textit{radius}) & $w_0 = 0.05$[m] ($5$[cm])\\
		\textit{Initial spot curvature} & $R_0 = \infty$ (collimated beam)\\
		\textit{Gaussian beam parameter} & $\Lambda = \frac{2L}{kw_L^2}$; $L$ measured in [m]\\
		\textit{$q$ parameter} & $q = 0.74\sigma_R^2Q_m^{\frac{1}{6}}$, $Q_m=35.05\frac{L}{k\times l_0^2}$; $L$, $l_0$ measured in [m]\\
		\textit{Extinction factor at sea level} & $\alpha_0 = 5\times 10^{-6}$[m$^{-1}$] \\
		\textit{Altitude} & $h_0 = 30$[m]\\
		\textit{Turbulence inner scale} & $l_0 = 0.001$[m] ($1$[mm])\\
		\textit{Phase structure function} & $D_\phi = \left(6.88\frac{r}{r_0}\right)^{\frac{5}{3}}=\sigma^2_\phi$\\
		\textit{Fried's turbulent parameter}& $r_0=\left(0.423C^2_nk^2L\right)^{-\frac{3}{5}}$; $L$ measured in [m] \\
		\textit{Interferometer arms separation} &  $r$ ($r>r_0$ \textit{or} $r=r_0$ \textit{or} $r<r_0$)
	\end{tabular}
\end{table}

\begin{figure}[!hbt]
	\centering
	\includegraphics[scale=0.4]{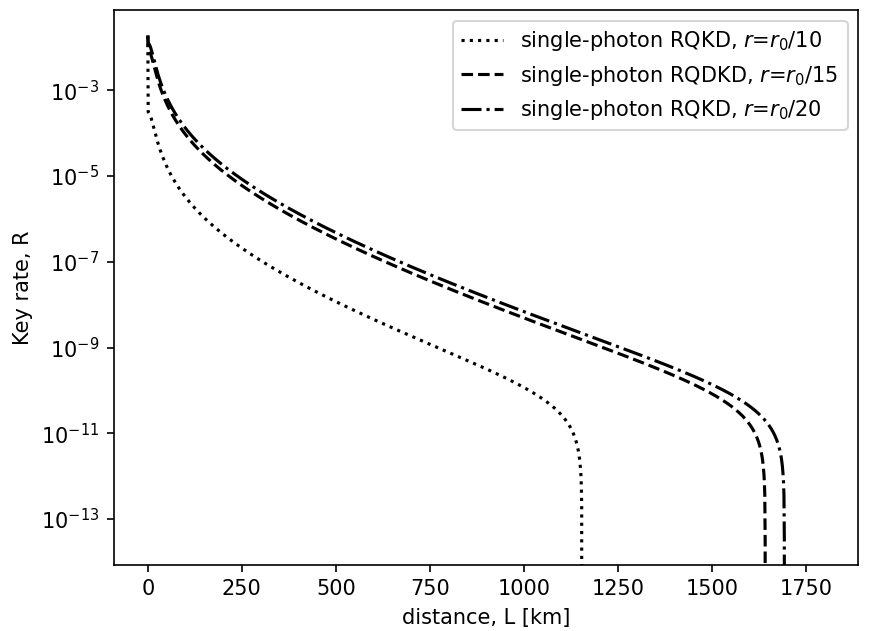}
	\caption{\label{single-photon-figure} Key\textendash rate of a single\textendash photon RQKD over a turbulent FSO channel. Three scenarios are considered: \textbf{(i)} $r=\frac{r_0}{10}$; \textbf{(ii)} $r=\frac{r_0}{15}$; \textbf{(iii)} $r=\frac{r_0}{20}$. Note that $r_0$ is the Fried's parameter (see Tab. \ref{table-parameters} for reference) and $r$ is the interferometer arms separation (see Fig. \ref{rqkd-fso}, Tab. \ref{table-parameters}).}
\end{figure}

\begin{table}[!hbt]
	\caption{\label{coherent-table-parameters} Parameters used in evaluating only the key rate of a coherent-state RQKD over a turbulent FSO. Values below are taken in accordance with Ref. \cite{lin-lutken}. }
	\begin{tabular}{c|c}
		\hline
		\textbf{Quantity} & \textbf{Value}/\textbf{Expression}\\
		\hline\hline
		\textit{Intensity} (\textit{Mode}) \textit{mismatch} & $V=0.95$\\
		\textit{Phase mismatch} & $\delta=\frac{\pi}{60}$\\
		\textit{Error correction inefficiency} & $f_{\text{EC}}=1.15$ \\
		\textit{Detector efficiency} & $\eta_{\text{det}}=0.5$ \\
		\textit{Dark count rate} & $p_{\text{dark}} = 10^{-10}$\\
		
	\end{tabular}
\end{table}
To evaluate the key rate of a single-photon RQKD protocol, we directly make use of Eq. \eqref{rate-loss} given the parameters introduced in Tab. \ref{table-parameters}. Take into consideration Appendix \ref{math-single-photon}, where the phase turbulence is incorporated into the mathematical description of the single-photon RQKD. Three scenarios are considered: \textbf{(i)} $r=\frac{r_0}{10}$; \textbf{(ii)} $r=\frac{r_0}{15}$; and \textbf{(iii)} $r=\frac{r_0}{20}$. The results are presented in Fig. \ref{single-photon-figure}. \\
%\fbox{\color{red}\textit{Remark:}} The original RQKD has zero key rate for $r>r_0$ and $r=r_0$.\\
%\fbox{\color{red}\textit{Remark:}} The interferometer arms separation $r$ is required to take a relatively significant value\textemdash ... \cite{kravtsov}. \\
%Based on the above remarks, we choose $r$ to be as close as possible to $r_0$ while ensuring that the original RQKD achieves a non\textendash zero key rate. \\
%\indent The key rates for the three scenarios are presented in Fig. \ref{comparison}. Note that the key rate of the proposed protocol (Polarization\textendash based RQKD) is independent of $r$ (a single plot for all considered scenarios). It is evident from the figure that
%\begin{enumerate}
%	\item[\textemdash] the proposed protocol achieves longer communication distances in all considered scenarios; 
%	\item[\textemdash] the proposed protocol achieves higher key rates than the original RQKD in scenarios \textbf{(i)} and \textbf{(ii)} ;
%	\item[\textemdash] the proposed protocol achieves slightly lower key rates than the original RQKD in scenario \textbf{(iii)}.
%\end{enumerate}
\begin{figure}[!hbt]
	\centering
	\includegraphics[scale=0.4]{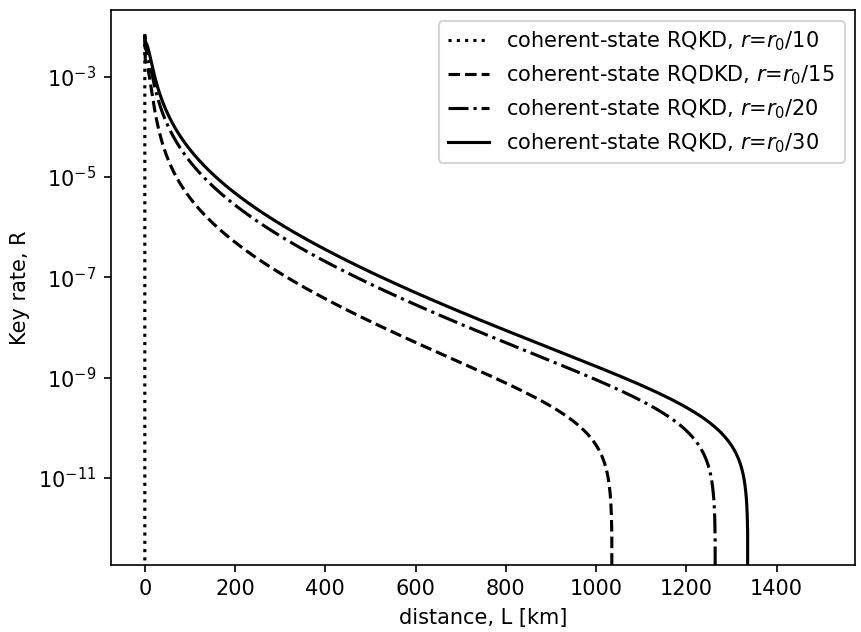}
	\caption{\label{coherent-state-figure} Key\textendash rate of a coherent\textendash state RQKD over a turbulent FSO channel. Four scenarios are considered: \textbf{(i)} $r=\frac{r_0}{10}$; \textbf{(ii)} $r=\frac{r_0}{15}$; \textbf{(iii)} $r=\frac{r_0}{20}$; \textbf{(iv)} $r=\frac{r_0}{30}$. Note that $r_0$ is the Fried's parameter (see Tab. \ref{table-parameters} for reference) and $r$ is the interferometer arms separation (see Fig. \ref{coherent-rqkd-fso}, Tab. \ref{table-parameters}). It is worth noting that the key rate vanishes for $L>0$ in case \textbf{(i)} (\textit{i.e.}, if $r\geq \frac{r_0}{10}$).}
\end{figure}
\subsection{Numerical results of coherent-state RQKD}\label{nr-coherent}
To determine the overall key rate of a coherent-state RQKD (see Eq. \eqref{overall-coherent-rate}), one is required to evaluate the Holevo information $\chi(K:E)_{\rho_{KYE}^{\gamma}}$, the error rate $e_-$ (correspondingly, $\delta^-_{\text{EC}}$), and the probability $p_{\gamma=-}$.\\
\textit{Evaluating $\chi(K:E)_{\rho_{KYE}^{\gamma}}$}\textemdash \\
\indent To evaluate $\chi(K:E)_{\rho_{KYE}^{\gamma}}$ (see Eq. \eqref{holevo}), we make use of the following computations. For the coherent-state RQKD, the Holevo information takes the form
\begin{equation}\label{holevo-certain}
	\begin{split}
		\chi(K:E)_{\rho_{KYE}^{-}} = S(\rho_E^-) - \sum\limits_k p_{k|-}\times S(\rho_E^{k,-}).
	\end{split}
\end{equation}
We adopt the approach of Ref. \cite{lin-lutken} to determine $\chi$ in the realistic (noisy) scenario. The Holevo information is evaluated using the same methodology as in the loss-only scenario, with the ideal POVM operators $F^\gamma$ replaced by the realistic POVM operators $F^\gamma_{\text{real}}$.\\
\textbf{Methodology to determine} $\chi(K:E)_{\rho_{KYE}^{-}}$ \textbf{in the loss-only scenario}\\
\indent In the loss-only scenario, $S(\rho^{k,-}_E) = 0$ because $\rho^{k,-}_E$ are pure states. The conditional state $\rho^-_E$ has the form
\begin{equation}
	\begin{split}
		\rho^-_E = \frac{1}{2}\left(\ketbra{\Theta^-_{0,0}}+\ketbra{\Theta^-_{1,1}}\right)
	\end{split}
\end{equation}
As mentioned in the protocol steps of Section \ref{coherent-rqkd}, outcome $\gamma=-$ in principle appears whenever \textit{Alice} and \textit{Bob} pick same phase shifts ($\varphi_a = \varphi_b$). Thus $\rho^-_E$ is characterized by the eigenvalues $\frac{1}{2}(1\pm \abs{\braket{\Theta^-_{0,0}}{\Theta^-_{1,1}}}$), as pointed out in Ref. \cite{lin-lutken}. The entropy $S(\rho^-_E)$ is then evaluated by $S(\rho^-_E)=H(\frac{1-\abs{\braket{\Theta^-_{0,0}}{\Theta^-_{1,1}}}}{2})$. Therefore $\chi(K:E)_{\rho_{KYE}^{-}}$ has the form 
\begin{equation}\label{chi-loss}
	\begin{split}
		\chi(K:E)_{\rho_{KYE}^{-}} &= S(\rho_E^-) - \sum\limits_k p_{k|-}\times S(\rho_E^{k,-})\\
		& = S(\rho_E^-) = H(\frac{1-\abs{\braket{\Theta^-_{0,0}}{\Theta^-_{1,1}}}}{2}).
	\end{split}
\end{equation}
in the loss-only scenario. \\
\textbf{Incorporating realistic POVM operators}\\
In the realistic scenario (FSO channel), the conditional state of \textit{Eve} $\ket{\Theta_{k,y}^{\gamma=-}}$ takes the form
\begin{equation}
	\begin{split}
		\ket{\tilde{\Theta}_{k,y}^{-}} = \frac{\sqrt{F^-_{\text{real}}}\ket{\varphi_k,\varphi_y}}{\sqrt{\bra{\varphi_k,\varphi_y}F^-_{\text{real}}\ket{\varphi_k,\varphi_y}}}.
	\end{split}
\end{equation}
where the realistic-scenario POVM operator $F^-_{\text{real}}$ is introduced in Appendix \ref{}. Note that 
\begin{equation}
	\begin{split}
		\braket{\tilde{\Theta}^-_{0,0}}{\tilde{\Theta}^-_{1,1}} &= \frac{\bra{\varphi_0,\varphi_0}F^-_{\text{real}} \ket{\varphi_1,\varphi_1}}{\sqrt{\bra{\varphi_0,\varphi_0}F^-_{\text{real}} \ket{\varphi_0,\varphi_0}\bra{\varphi_1,\varphi_1}F^-_{\text{real}} \ket{\varphi_1,\varphi_1}}}\\
		&=
		\frac{\bra{\sqrt{\mu},\sqrt{\mu}}F^-_{\text{real}} \ket{-\sqrt{\mu},-\sqrt{\mu}}}{\sqrt{\bra{\sqrt{\mu},\sqrt{\mu}}F^-_{\text{real}} \ket{\sqrt{\mu},\sqrt{\mu}}\bra{-\sqrt{\mu},-\sqrt{\mu}}F^-_{\text{real}} \ket{-\sqrt{\mu},-\sqrt{\mu}}}}
	\end{split}
\end{equation}
Then invoking expression \eqref{chi-loss}, one yields
\begin{equation}
	\begin{split}
		\chi(K:E)_{\rho_{KYE}^{-},\text{\textbf{real}}} &=  H\left(\frac{1-\abs{\braket{\tilde{\Theta}^-_{0,0}}{\tilde{\Theta}^-_{1,1}}}}{2}\right).
	\end{split}
\end{equation}
\textit{Evaluating $e_-$}\textemdash \\
\indent In the coherent-state protocol, $e_-$ has the form \cite{lin-lutken}
\begin{equation}
	\begin{split}
		e_- = \frac{\zeta_2 - (1-p_{\text{dark}})\zeta_1\zeta_2}{\zeta_1+\zeta_2-2(1-p_{\text{dark}})\zeta_1\zeta_2} 
	\end{split}
\end{equation}
where 
\begin{equation}
	\begin{split}
		\zeta_1 &= \mathrm{e}^{-\sqrt{\eta}\mu[1-\sqrt{V}\langle\cos(\delta+\Delta\phi_{\text{turb}})\rangle]}\\
		\zeta_2 &= \mathrm{e}^{-\sqrt{\eta}\mu[1+\sqrt{V}\langle\cos(\delta+\Delta\phi_{\text{turb}})\rangle]}
	\end{split}
\end{equation} 
Note that an average is taken over the cosine due to incorporating \textit{phase turbulence} into the phase misalignment. We assume that $\Delta\phi_{\text{turb}} \sim \mathcal{N}(\Delta\overline{\phi},2\sigma^2_{\phi})$ (as in the single-photon regime, see \eqref{turb-gauss}). Suppose that $\Delta\overline{\phi} = 0$. Then, based on expression \eqref{cos-differ}, we have
\begin{equation}\label{cos-turb}
	\begin{split}
		\langle\cos(\delta+\Delta\phi_{\text{turb}})\rangle = \mathrm{e}^{-\frac{2\sigma^2_{\phi}}{2}}\cos(\delta) = \mathrm{e}^{-\sigma^2_{\phi}}\cos(\delta)
	\end{split}
\end{equation}
where 
\begin{equation}
	\begin{split}
		\delta+\Delta\phi_{\text{turb}} &= \delta + \left(\Delta\overline{\phi} + 2\sigma^2_{\phi}\right) \\
		&= \delta + 2\sigma^2_{\phi}
	\end{split}
\end{equation}
is considered a Gaussian variable ($\delta+\Delta\phi_{\text{turb}} \sim \mathcal{N}(\delta,2\sigma^2_{\phi})$). Therefore $\zeta_1$ and $\zeta_2$ take the form
\begin{equation}
	\begin{split}
		\zeta_1 &= \mathrm{e}^{-\sqrt{\eta}\mu[1-\sqrt{V}\mathrm{e}^{-\sigma^2_{\phi}}\cos(\delta)]},\\
		\zeta_2 &= \mathrm{e}^{-\sqrt{\eta}\mu[1+\sqrt{V}\mathrm{e}^{-\sigma^2_{\phi}}\cos(\delta)]}.
	\end{split}
\end{equation}
The last expressions imply that the phase turbulence leads to modifying the visibility. Knowing $e_-$, one is capable of determining $\delta_{\text{EC}}$ by \eqref{ec}. 
\textit{Evaluating $p_{\gamma=-}$}\textemdash \\
The probability $p_{\gamma=-}$ is given by
\begin{equation}
	\begin{split}
		p_{\gamma=-} = \sum\limits_{k,y} p_{k,y}p_{-|k,y}.
	\end{split}
\end{equation}
Since \textit{Alice} and \textit{Bob} choose \textit{their} coherent states (signals) $\ket{\varphi_x}$ and $\ket{\varphi_y}$ in a random manner, then $p_{k,y} = \frac{1}{4}$ ($\forall k,y$). According to Ref. \cite{lin-lutken}, the conditional probabilities $p_{-|k,y}$ are given by
\begin{equation}
	\begin{split}
		p_{-|k,y} = \bra{\varphi_k,\varphi_y}F^-_{\text{real}}\ket{\varphi_k,\varphi_y}.
	\end{split}
\end{equation}
Hence
\begin{equation}
	\begin{split}
		p_{\gamma=-} = &\frac{1}{4}\Big[\bra{\varphi_0,\varphi_0}F^-_{\text{real}}\ket{\varphi_0,\varphi_0}+\bra{\varphi_0,\varphi_1}F^-_{\text{real}}\ket{\varphi_0,\varphi_1}\\
		&+\bra{\varphi_1,\varphi_0}F^-_{\text{real}}\ket{\varphi_1,\varphi_0} + \bra{\varphi_1,\varphi_1}F^-_{\text{real}}\ket{\varphi_1,\varphi_1}\Big]
	\end{split}
\end{equation}
which can be rewritten as
\begin{equation}
	\begin{split}
		p_{\gamma=-} = &\frac{1}{4}\Big[\bra{\sqrt{\mu},\sqrt{\mu}}F^-_{\text{real}}\ket{\sqrt{\mu},\sqrt{\mu}}+\bra{\sqrt{\mu},-\sqrt{\mu}}F^-_{\text{real}}\ket{\sqrt{\mu},-\sqrt{\mu}}\\
		&+\bra{-\sqrt{\mu},\sqrt{\mu}}F^-_{\text{real}}\ket{-\sqrt{\mu},\sqrt{\mu}} + \bra{-\sqrt{\mu},-\sqrt{\mu}}F^-_{\text{real}}\ket{-\sqrt{\mu},-\sqrt{\mu}}\Big]
	\end{split}
\end{equation}

\section{Conclusions}\label{conclusions}
In this paper, two realizations of relativistic quantum key distribution (RQKD) over turbulent free-space optical (FSO) channels have been investigated: a single-photon implementation and a coherent-state implementation. A mathematical model describing the influence of atmospheric turbulence on the interferometric visibility and, consequently, on the secret-key generation process has been developed. The resulting key-rate expressions were evaluated numerically under realistic atmospheric and experimental parameters.\\
\indent The numerical results demonstrate that the communication performance strongly depends on the ratio between the interferometer-arm separation and Fried's parameter, $\frac{r}{r_0}$, which quantifies the strength of turbulence-induced phase fluctuations. As the separation between the interferometer arms increases relative to $r_0$, atmospheric turbulence progressively suppresses interference visibility, leading to higher quantum bit error rates and lower secret-key rates. \\
\indent For the single-photon protocol, all investigated scenarios produce non-zero secret-key rates, although the achievable communication distance decreases as the turbulence becomes stronger. The protocol therefore exhibits a relatively robust behavior under moderate atmospheric disturbances.\\
\indent For the coherent-state implementation, the effect of turbulence is significantly more pronounced. The obtained results reveal the existence of a critical interferometer-arm separation beyond which secure key generation rapidly becomes impossible. In particular, for $r=\frac{r_0}{10}$, the secret-key rate vanishes for any non-zero transmission distance, whereas reducing the arm separation to $r=\frac{r_0}{15}$, $r=\frac{r_0}{20}$, and $r=\frac{r_0}{30}$ progressively restores secure operation and extends the achievable communication range. These results clearly illustrate the importance of minimizing the spatial separation between the interferometer arms (the temporal separation $\Delta T$) in practical coherent-state RQKD systems.\\
\indent Overall, the presented analysis establishes a quantitative relationship between atmospheric turbulence and the achievable secret-key rate of relativistic QKD systems. The derived model provides a practical framework for predicting protocol performance under realistic free-space conditions and identifies the interferometer geometry as one of the principal design parameters governing secure communication. The obtained results may therefore serve as useful guidelines for the implementation and optimization of future free-space relativistic QKD systems operating in turbulent atmospheric environments.

\appendix
\section{Mathematical description}\label{math-single-photon}
In this section, a mathematical description of the a single\textendash photon RQKD setup (Fig. \ref{rqkd-fso}) is put forward\textemdash the state of a single-photon system is mathematically tracked throughout the quantum phase of RQKD protocol.\\
\indent Initially \textit{Alice} prepares a single-photon $q_k$ (certain $k$) system: $a^\dagger\ket{0}$. Next the quantum system is incident on a beam splitter (\textit{balanced} beam splitter). The beam splitter puts $q_k$ into a spatial superposition:
\begin{equation*}
	\begin{split}
		a^\dagger\ket{0} \rightarrow \frac{1}{\sqrt{2}}\left(u^\dagger+l^\dagger\right)\ket{0}.
	\end{split}
\end{equation*}
Next \textit{Alice} induces at random a \textit{phase shift} $\phi_a$ ($\phi_a\in\{0,\frac{\pi}{2}\}$)\footnote{$\phi_a = 0$ corresponds to not applying a phase shift operator $\phi_a$ to the optical mode of concern.} to the upper\textendash arm mode. This results in a mixture
\begin{equation*}
	\begin{split}
		\rho = \frac{1}{2}\ketbra{\psi_0}+\frac{1}{2}\ketbra{\psi_{\frac{\pi}{2}}},
	\end{split}
\end{equation*}
where
\begin{equation*}
	\begin{split}
		\ket{\psi_0} &= \frac{1}{\sqrt{2}}\left(\mathrm{e}^{\mathrm{i}\phi_a}u^\dagger+l^\dagger\right)\ket{0}= \frac{1}{\sqrt{2}}\left(\mathrm{e}^{\mathrm{i}0}u^\dagger+l^\dagger\right)\ket{0},\\
		\ket{\psi_{\frac{\pi}{2}}} &= \frac{1}{\sqrt{2}}\left(\mathrm{e}^{\mathrm{i}\phi_a}u^\dagger+l^\dagger\right)\ket{0}= \frac{1}{\sqrt{2}}\left(\mathrm{e}^{\mathrm{i}\frac{\pi}{2}}u^\dagger+l^\dagger\right)\ket{0}.
	\end{split}
\end{equation*}
Then the upper\textendash arm mode undergoes a reflection off a mirror (a phase of $\pi$ is induced in the upper\textendash arm mode):
\begin{equation*}
	\begin{split}
		\ket{\psi_0} &= \frac{1}{\sqrt{2}}\left(-\mathrm{e}^{\mathrm{i}0}u^\dagger+l^\dagger\right)\ket{0},\\
		\ket{\psi_{\frac{\pi}{2}}} &=  \frac{1}{\sqrt{2}}\left(-\mathrm{e}^{\mathrm{i}\frac{\pi}{2}}u^\dagger+l^\dagger\right)\ket{0}.
	\end{split}
\end{equation*}
Both upper\textendash$\;$ and lower\textendash arm modes independently pick phase shifts out of the quantum channel ($\phi_u$ and $\phi_l$ phase shifts):
\begin{equation*}
	\begin{split}
		\ket{\psi_0} &= \frac{1}{\sqrt{2}}\left(-\mathrm{e}^{\mathrm{i}0}\mathrm{e}^{\mathrm{i}\phi_u}u^\dagger+\mathrm{e}^{\mathrm{i}\phi_l}l^\dagger\right)\ket{0},\\
		\ket{\psi_{\frac{\pi}{2}}} &=  \frac{1}{\sqrt{2}}\left(-\mathrm{e}^{\mathrm{i}\frac{\pi}{2}}\mathrm{e}^{\mathrm{i}\phi_u}u^\dagger+\mathrm{e}^{\mathrm{i}\phi_l}l^\dagger\right)\ket{0}.
	\end{split}
\end{equation*}
Next the lower\textendash arm mode undergoes a reflection off a mirror (a phase of $\pi$ is induced in the upper\textendash arm mode):
\begin{equation*}
	\begin{split}
		\ket{\psi_0} &= \frac{1}{\sqrt{2}}\left(-\mathrm{e}^{\mathrm{i}0}\mathrm{e}^{\mathrm{i}\phi_u}u^\dagger-\mathrm{e}^{\mathrm{i}\phi_l}l^\dagger\right)\ket{0}= - \frac{1}{\sqrt{2}}\left(\mathrm{e}^{\mathrm{i}0}\mathrm{e}^{\mathrm{i}\phi_u}u^\dagger+\mathrm{e}^{\mathrm{i}\phi_l}l^\dagger\right)\ket{0},\\
		\ket{\psi_{\frac{\pi}{2}}} &=  \frac{1}{\sqrt{2}}\left(-\mathrm{e}^{\mathrm{i}\frac{\pi}{2}}\mathrm{e}^{\mathrm{i}\phi_u}u^\dagger-\mathrm{e}^{\mathrm{i}\phi_l}l^\dagger\right)\ket{0} = - \frac{1}{\sqrt{2}}\left(\mathrm{e}^{\mathrm{i}\frac{\pi}{2}}\mathrm{e}^{\mathrm{i}\phi_u}u^\dagger+\mathrm{e}^{\mathrm{i}\phi_l}l^\dagger\right)\ket{0}.
	\end{split}
\end{equation*}
\textit{Bob} induces at random a \textit{phase shift} $\phi_b$ ($\phi_b\in\{0,\frac{\pi}{2}\}$) to the lower\textendash arm mode. This results in a mixture
\begin{equation*}
	\begin{split}
		\rho' = \frac{1}{4}\ketbra{\psi_{0,0}} + \frac{1}{4}\ketbra{\psi_{0,\frac{\pi}{2}}} + \frac{1}{4}\ketbra{\psi_{\frac{\pi}{2},0}} + \frac{1}{4}\ketbra{\psi_{\frac{\pi}{2},\frac{\pi}{2}}},
	\end{split}
\end{equation*}
where 
\begin{equation*}
	\begin{split}
		\ket{\psi_{0,0}} & = - \frac{1}{\sqrt{2}}\left(\mathrm{e}^{\mathrm{i}0}\mathrm{e}^{\mathrm{i}\phi_u}u^\dagger+\mathrm{e}^{\mathrm{i}\phi_l}\mathrm{e}^{\mathrm{i}0}l^\dagger\right)\ket{0},\\
		\ket{\psi_{0,\frac{\pi}{2}}} & = - \frac{1}{\sqrt{2}}\left(\mathrm{e}^{\mathrm{i}0}\mathrm{e}^{\mathrm{i}\phi_u}u^\dagger+\mathrm{e}^{\mathrm{i}\phi_l}\mathrm{e}^{\mathrm{i}\frac{\pi}{2}}l^\dagger\right)\ket{0},\\
		\ket{\psi_{\frac{\pi}{2},0}} & = - \frac{1}{\sqrt{2}}\left(\mathrm{e}^{\mathrm{i}\frac{\pi}{2}}\mathrm{e}^{\mathrm{i}\phi_u}u^\dagger+\mathrm{e}^{\mathrm{i}\phi_l}\mathrm{e}^{\mathrm{i}0}l^\dagger\right)\ket{0},\\
		\ket{\psi_{\frac{\pi}{2},\frac{\pi}{2}}} & = - \frac{1}{\sqrt{2}}\left(\mathrm{e}^{\mathrm{i}\frac{\pi}{2}}\mathrm{e}^{\mathrm{i}\phi_u}u^\dagger+\mathrm{e}^{\mathrm{i}\phi_l}\mathrm{e}^{\mathrm{i}\frac{\pi}{2}}l^\dagger\right)\ket{0}.
	\end{split}
\end{equation*}
Note that the first subscript of the ket state reflects the phase induced by \textit{Alice} and the second one reflects the phase induced by \textit{Bob}. The last ket states can be rewritten as
\begin{equation*}
	\begin{split}
		\ket{\psi_{0,0}} & = -\mathrm{e}^{\mathrm{i}(0+\phi_u)} \frac{1}{\sqrt{2}}\left(u^\dagger+\mathrm{e}^{\mathrm{i}[(\phi_l+0)-(0+\phi_u)]}l^\dagger\right)\ket{0},\\
		\ket{\psi_{0,\frac{\pi}{2}}} & = -\mathrm{e}^{\mathrm{i}(0+\phi_u)} \frac{1}{\sqrt{2}}\left(u^\dagger+\mathrm{e}^{\mathrm{i}[(\phi_l+\frac{\pi}{2})-(0+\phi_u)]}l^\dagger\right)\ket{0},\\
		\ket{\psi_{\frac{\pi}{2},0}} & = -\mathrm{e}^{\mathrm{i}(\frac{\pi}{2}+\phi_u)} \frac{1}{\sqrt{2}}\left(u^\dagger+\mathrm{e}^{\mathrm{i}[(\phi_l+0)-(\frac{\pi}{2}+\phi_u)]}l^\dagger\right)\ket{0},\\
		\ket{\psi_{\frac{\pi}{2},\frac{\pi}{2}}} & = -\mathrm{e}^{\mathrm{i}(\frac{\pi}{2}+\phi_u)} \frac{1}{\sqrt{2}}\left(u^\dagger+\mathrm{e}^{\mathrm{i}[(\phi_l+\frac{\pi}{2}) - (\frac{\pi}{2}+\phi_u)]}l^\dagger\right)\ket{0}.
	\end{split}
\end{equation*}
Recall that the global phases above have no effect on $\rho'$. Next upper\textendash$\;$ and lower\textendash arm modes interfere at the second beam splitter. At the beam splitter the following transformations take place
\begin{equation*}
	\begin{split}
		u^\dagger &\rightarrow \frac{1}{\sqrt{2}}\left(c^\dagger + d^\dagger\right),\\
		l^\dagger &\rightarrow \frac{1}{\sqrt{2}}\left(c^\dagger - d^\dagger\right),
	\end{split}
\end{equation*}
where $c^\dagger$ and $d^\dagger$ denote the output modes of the second beam splitter. Therefore
\begin{equation}\label{rqkd-final-state}
	\begin{split}
		\rho' \rightarrow \rho'' = \frac{1}{4}\ketbra{\tilde{\psi}_{0,0}} + \frac{1}{4}\ketbra{\tilde{\psi}_{0,\frac{\pi}{2}}} + \frac{1}{4}\ketbra{\tilde{\psi}_{\frac{\pi}{2},0}} + \frac{1}{4}\ketbra{\tilde{\psi}_{\frac{\pi}{2},\frac{\pi}{2}}}.
	\end{split}
\end{equation}
where
\begin{equation*}
	\begin{split}
		\ket{\tilde{\psi}_{0,0}} & = -\mathrm{e}^{\mathrm{i}(0+\phi_u)} \frac{1}{\sqrt{2}}\left(\frac{1}{\sqrt{2}}\left(c^\dagger + d^\dagger\right)+\mathrm{e}^{\mathrm{i}[(\phi_l+0)-(0+\phi_u)]}\frac{1}{\sqrt{2}}\left(c^\dagger - d^\dagger\right)\right)\ket{0},\\
		\ket{\tilde{\psi}_{0,\frac{\pi}{2}}} & = -\mathrm{e}^{\mathrm{i}(0+\phi_u)} \frac{1}{\sqrt{2}}\left(\frac{1}{\sqrt{2}}\left(c^\dagger + d^\dagger\right)+\mathrm{e}^{\mathrm{i}[(\phi_l+\frac{\pi}{2})-(0+\phi_u)]}\frac{1}{\sqrt{2}}\left(c^\dagger - d^\dagger\right)\right)\ket{0},\\
		\ket{\tilde{\psi}_{\frac{\pi}{2},0}} & = -\mathrm{e}^{\mathrm{i}(\frac{\pi}{2}+\phi_u)} \frac{1}{\sqrt{2}}\left(\frac{1}{\sqrt{2}}\left(c^\dagger + d^\dagger\right)+\mathrm{e}^{\mathrm{i}[(\phi_l+0)-(\frac{\pi}{2}+\phi_u)]}\frac{1}{\sqrt{2}}\left(c^\dagger - d^\dagger\right)\right)\ket{0},\\
		\ket{\tilde{\psi}_{\frac{\pi}{2},\frac{\pi}{2}}} & = -\mathrm{e}^{\mathrm{i}(\frac{\pi}{2}+\phi_u)} \frac{1}{\sqrt{2}}\left(\frac{1}{\sqrt{2}}\left(c^\dagger + d^\dagger\right)+\mathrm{e}^{\mathrm{i}[(\phi_l+\frac{\pi}{2}) - (\frac{\pi}{2}+\phi_u)]}\frac{1}{\sqrt{2}}\left(c^\dagger - d^\dagger\right)\right)\ket{0}.
	\end{split}
\end{equation*}
A detector is placed only at the $d$\textendash output of BS\textemdash a detection click corresponds to a key\textendash generation event. To describe that a quantum system is incident on the detector, we apply the following projector
\begin{equation}\label{projector-d}
	\begin{split}
		\Pi_d = d^\dagger\ketbra{0}d.
	\end{split}
\end{equation}
That is,
\begin{equation}\label{rqkd-output}
	\begin{split}
		\Pi_d \rho'' \Pi^\dagger_d &= \frac{1}{4}\Pi_d \ketbra{\tilde{\psi}_{0,0}}\Pi^\dagger_d + \frac{1}{4}\Pi_d \ketbra{\tilde{\psi}_{0,\frac{\pi}{2}}}\Pi^\dagger_d \\
		&+ \frac{1}{4}\Pi_d \ketbra{\tilde{\psi}_{\frac{\pi}{2},0}}\Pi^\dagger_d + \frac{1}{4}\Pi_d \ketbra{\tilde{\psi}_{\frac{\pi}{2},\frac{\pi}{2}}}\Pi^\dagger_d \\
		&=\frac{1}{4}\Big[ \frac{1}{2}(1-\cos(\phi_l-\phi_u))d^\dagger\ketbra{0}d + \frac{1}{2}(1+\sin(\phi_l-\phi_u))d^\dagger\ketbra{0}d \\
		&+ \frac{1}{2}(1-\sin(\phi_l-\phi_u))d^\dagger\ketbra{0}d + \frac{1}{2}(1-\cos(\phi_l-\phi_u))d^\dagger\ketbra{0}d\Big].
	\end{split}
\end{equation}
Hence the probability for a click to take place is
\begin{equation}\label{prob-rqkd}
	\begin{split}
		p_{\text{det}} = \Tr(\Pi_d\rho''\Pi^\dagger_d) &= \frac{1}{4}\Big[ \frac{1}{2}(1-\cos(\phi_l-\phi_u)) + \frac{1}{2}(1+\sin(\phi_l-\phi_u)) \\
		&+ \frac{1}{2}(1-\sin(\phi_l-\phi_u)) + \frac{1}{2}(1-\cos(\phi_l-\phi_u))\Big]\\
		&= \frac{1}{2}-\frac{1}{4}\cos(\phi_l-\phi_u).
	\end{split}
\end{equation}
Note that $p_{\text{det}}$ is a function of both $\phi_l$ and $\phi_u$ ($p_{\text{det}}(\phi_l,\phi_u)$).\\
\underline{\textbf{RQKD over FSO channel} with \textit{phase turbulence}}. \\
\indent The phase shifts $\phi_u$ and $\phi_l$ are modeled as random variables with standard deviation $\sigma_{\phi}$ in a turbulent free\textendash space optical (FSO) channel \cite{}:
\begin{equation*}
	\begin{split}
		\phi_u &= \overline{\phi}_u + \delta\phi_u, \; \phi_u\sim\mathcal{N}(\overline{\phi}_u,\sigma_\phi^2)	\\
		\phi_l &= \overline{\phi}_l + \delta\phi_l, \; \phi_l\sim\mathcal{N}(\overline{\phi}_l,\sigma_\phi^2) 
	\end{split}
\end{equation*}
where $\overline{\phi}_u$ ($\overline{\phi}_l$) is the ideal value of $\phi_u$ ($\phi_l$), $\delta\phi_u$ ($\delta\phi_l$) is the fluctuation of $\phi_u$ ($\phi_l$), and $\mathcal{N}(\text{mean},\text{variance})$ denotes a normal distribution. Note that $\overline{\phi}_u = \overline{\phi}_l = 0$ [rad]. Thus
\begin{equation*}
	\begin{split}
		\phi_u = \delta\phi_u  \sim\mathcal{N}(0,\sigma_\phi^2),\\
		\phi_l = \delta\phi_l \sim\mathcal{N}(0,\sigma_\phi^2).
	\end{split}
\end{equation*}
The above lines imply that the phase difference $\phi_l - \phi_u$ transforms into
\begin{equation*}
	\begin{split}
		\phi_l - \phi_u = \overline{\phi}_l + \delta\phi_l - \left(\overline{\phi}_u + \delta\phi_u\right) = (\overline{\phi}_l-\overline{\phi}_u) + (\delta\phi_l-\delta\phi_u),
	\end{split}
\end{equation*}
that is,
\begin{equation*}
	\begin{split}
		\Delta\phi = \Delta\overline{\phi} + \delta\Delta\phi.
	\end{split}
\end{equation*}
Since we assume that the fluctuations of $\phi_u$ and $\phi_l$ are independent of each other, then
\begin{equation*}
	\begin{split}
		\mathrm{Var}\left(\delta\Delta\phi\right) = \mathrm{Var}\left(\delta\phi_l-\delta\phi_u\right).
	\end{split}
\end{equation*}
Making use of $\mathrm{Var}(A-B) = \mathrm{Var}(A) + \mathrm{Var}(B) - 2\mathrm{Cov}(A,B)$ and that $\mathrm{Cov}(A,B)=0$ if $A$ and $B$ are independent (uncorrelated) \cite{}, we get
\begin{equation*}
	\begin{split}
		\mathrm{Var}\left(\delta\Delta\phi\right) = \mathrm{Var}(\delta\phi_l) + \mathrm{Var}(\delta\phi_u) = \sigma_\phi^2 + \sigma_\phi^2 = 2\sigma_\phi^2.
	\end{split}
\end{equation*}
Hence
\begin{equation}\label{turb-gauss}
	\begin{split}
		\Delta\phi\sim\mathcal{N}\left(\Delta\overline{\phi},2\sigma_\phi^2\right).
	\end{split}
\end{equation}
Since the probability of a detector click $p'_{\text{det}}$ in a RQKD is a function of both $\phi_u$ and $\phi_l$ (see Eq. \eqref{prob-rqkd}), we evaluate its expectation value for the sake of determining the key rate of RQKD over FSO. Its expectation value takes the form
\begin{equation}\label{expected-prob}
	\begin{split}
		\langle p'_{\text{det}} \rangle = \langle \frac{1}{2}-\frac{1}{4}\cos(\phi_l-\phi_u) \rangle = \frac{1}{2}-\frac{1}{4}\langle\cos(\Delta\phi)\rangle. 
	\end{split}
\end{equation}
Because $\Delta\phi\sim\mathcal{N}\left(\Delta\overline{\phi},2\sigma_\phi^2\right)$, we obtain \cite{}
\begin{equation}\label{cos-differ}
	\begin{split}
		\langle\cos(\Delta\phi)\rangle &= \mathrm{e}^{-\frac{2\sigma_\phi^2}{2}}\cos(\Delta\overline{\phi}) \\
		&= \mathrm{e}^{-\sigma_\phi^2}\cos(\Delta\overline{\phi}).
	\end{split}
\end{equation}
That is,
\begin{equation*}
	\begin{split}
		\langle p'_{\text{det}} \rangle &= \frac{1}{2}-\frac{1}{4}\mathrm{e}^{-\sigma_\phi^2}\cos(\Delta\overline{\phi}) \\
		&= \frac{1}{2}-\frac{1}{4}\mathrm{e}^{-\sigma_\phi^2}
	\end{split}
\end{equation*}
taking into account that $\Delta\overline{\phi} = \overline{\phi}_l-\overline{\phi}_u = 0 - 0 = 0$.\\
\indent \textit{Error rate}: In the RQKD framework, the error rate due to \textit{phase turbulence} is evaluated by observing the probability of registering a click at the detector given $\phi_a=\phi_b$. Therefore we examine 
\begin{equation*}
	\begin{split}
		\rho'' = \frac{1}{4}\ketbra{\tilde{\psi}_{0,0}} + \frac{1}{4}\ketbra{\tilde{\psi}_{0,\frac{\pi}{2}}} + \frac{1}{4}\ketbra{\tilde{\psi}_{\frac{\pi}{2},0}} + \frac{1}{4}\ketbra{\tilde{\psi}_{\frac{\pi}{2},\frac{\pi}{2}}}.
	\end{split}
\end{equation*}
as projector $\Pi_{0,0}+\Pi_{\frac{\pi}{2},\frac{\pi}{2}}$ is applied [\textit{Note}: $\rho''$ is introduced in Eq. \eqref{rqkd-final-state}]. That is,
\begin{equation*}
	\begin{split}
		\left(\Pi_{0,0}+\Pi_{\frac{\pi}{2},\frac{\pi}{2}}\right)\rho''(\Pi^\dagger_{0,0}+\Pi^\dagger_{\frac{\pi}{2},\frac{\pi}{2}}) = \rho''_{\phi_a=\phi_b},
	\end{split}
\end{equation*}
where
\begin{equation*}
	\begin{split}
		\Pi_{0,0} &= \ketbra{\tilde{\psi}_{0,0}},\\
		\Pi_{\frac{\pi}{2},\frac{\pi}{2}} &= \ketbra{\tilde{\psi}_{\frac{\pi}{2},\frac{\pi}{2}}}.
	\end{split}
\end{equation*}
This results in
\begin{equation*}
	\begin{split}
		\rho''_{\phi_a=\phi_b} = \frac{1}{4}\ketbra{\tilde{\psi}_{0,0}} + \frac{1}{4}\ketbra{\tilde{\psi}_{\frac{\pi}{2},\frac{\pi}{2}}}.
	\end{split}
\end{equation*}
Now we find the probability of a detector click. This is achieved by applying $\Pi_d$ (see Eq. \eqref{projector-d}) to $\rho''_{\phi_a=\phi_b}$:
\begin{equation*}
	\begin{split}
		\Pi_d\rho''_{\phi_a=\phi_b}\Pi^\dagger_d = \frac{1}{4}\Big[ \frac{1}{2}(1-\cos(\phi_l-\phi_u))d^\dagger\ketbra{0}d + \frac{1}{2}(1-\cos(\phi_l-\phi_u))d^\dagger\ketbra{0}d\Big].
	\end{split}
\end{equation*}
Hence, the probability of an erroneous click at the detector is given by
\begin{equation*}
	\begin{split}
		p'_{\text{error}} = \Tr(\Pi_d\rho''_{\phi_a=\phi_b}\Pi^\dagger_d) = \frac{1}{4}\left(1-\cos(\phi_l-\phi_u)\right).
	\end{split}
\end{equation*}
Given a phase turbulence, as proceeded in Eq. \eqref{expected-prob}, $p'_{\text{error}}$ takes the form
\begin{equation*}
	\begin{split}
		\langle p'_{\text{error}}\rangle = \langle \frac{1}{4}\left(1-\cos(\phi_l-\phi_u)\right) \rangle = \frac{1}{4}\left(1-\langle\cos(\Delta\phi)\rangle\right).
	\end{split}
\end{equation*} 
Taking into account the result of Eq. \eqref{cos-differ}, we get
\begin{equation*}
	\begin{split}
		\langle p'_{\text{error}}\rangle = \frac{1}{4} - \frac{1}{4}\mathrm{e}^{-\sigma_\phi^2}\cos(\Delta\overline{\phi}),
	\end{split}
\end{equation*}
which is reduced to
\begin{equation*}
	\begin{split}
		\langle p'_{\text{error}}\rangle = \frac{1}{4} - \frac{1}{4}\mathrm{e}^{-\sigma_\phi^2}
	\end{split}
\end{equation*}
since $\Delta\overline{\phi} = \overline{\phi}_l-\overline{\phi}_u = 0 - 0 = 0$.\\

\section{Purity of $\sigma_0$ and $\sigma_{\frac{\pi}{2}}$}\label{pure}
The \textit{purity} of a quantum state $\sigma=\ketbra{\psi}$ is given by \cite{}
\begin{equation}\label{purity}
	\begin{split}
		\Gamma = \Tr(\sigma^2).
	\end{split}
\end{equation}
This expression holds for \textit{normalized} states ($\braket{\psi}{\psi}=1$). Note that a state is pure \textit{if and only if} $\Gamma=1$. We first verify that $\ket{\psi_0}$ and $\ket{\psi_{\frac{\pi}{2}}}$ are normalized. Then we check if these states are \textit{pure} according to Eq. \eqref{purity}.\\
Since
\begin{equation}\label{commutators}
	\begin{split}
		\commutator{u}{u^\dagger} &= 1,\\
		\commutator{l}{l^\dagger} &= 1,\\
		\commutator{u}{l^\dagger} &= 0,\\
	\end{split}
\end{equation}
we get
\begin{equation}
	\begin{split}
		\braket{\psi_0}{\psi_0} &= \frac{1}{2}\bra{0}(\mathrm{e}^{-\mathrm{i}0}u+l)(\mathrm{e}^{\mathrm{i}0}u^\dagger+l^\dagger)\ket{0}\\
		&=\frac{1}{2}(\bra{0}uu^\dagger\ket{0} + \bra{0}ll^\dagger\ket{0})\\
		&= \frac{1}{2}(1+1)=1
	\end{split}
\end{equation}
and 
\begin{equation}
	\begin{split}
		\braket{\psi_{\frac{\pi}{2}}}{\psi_{\frac{\pi}{2}}} &= \frac{1}{2}\bra{0}(\mathrm{e}^{-\mathrm{i}\frac{\pi}{2}}u+l)(\mathrm{e}^{\mathrm{i}\frac{\pi}{2}}u^\dagger+l^\dagger)\ket{0}\\
		&=\frac{1}{2}(\bra{0}uu^\dagger\ket{0} + \bra{0}ll^\dagger\ket{0})\\
		&= \frac{1}{2}(1+1)=1.
	\end{split}
\end{equation}
Hence both $\ket{\psi_0}$ and $\ket{\psi_{\frac{\pi}{2}}}$ are normalized. We find that 
\begin{equation}
	\begin{split}
		\sigma_0^2 &= \ket{\psi_0}\braket{\psi_0}{\psi_0}\bra{\psi_0} = \ketbra{\psi_0} = \rho_0,\\
		\sigma_{\frac{\pi}{2}}^2 &= \ket{\psi_{\frac{\pi}{2}}}\braket{\psi_{\frac{\pi}{2}}}{\psi_{\frac{\pi}{2}}}\bra{\psi_{\frac{\pi}{2}}} = \ketbra{\psi_{\frac{\pi}{2}}} = \rho_{\frac{\pi}{2}}
	\end{split}
\end{equation}
\textit{i.e.}, $\sigma_0$ and $\sigma_{\frac{\pi}{2}}$ are \textit{idempotent} as well. Therefore 
\begin{equation}
	\begin{split}
		\Gamma_0 &= \Tr(\sigma_0^2) = \Tr(\sigma_0) = 1,\\
		\Gamma_{\frac{\pi}{2}} &= \Tr(\sigma_{\frac{\pi}{2}}^2) = \Tr(\sigma_{\frac{\pi}{2}}) = 1.\\
	\end{split}
\end{equation}
That is, $\sigma_0$ and $\sigma_{\frac{\pi}{2}}$ are pure.

\section{Coherent states $\ket{\varphi_x,\varphi_y}$}
As shown in \cite{lin-lutken}, a coherent state $\ket{\pm\sqrt{\mu}}$ can be written as
\begin{equation}
	\begin{split}
		\ket{\pm\sqrt{\mu}} = c_0\ket{e_0} \pm c_1\ket{e_1},
	\end{split}
\end{equation}
where states $\{\ket{e_i}\}$ form an orthonormal basis, and coefficients $\{c_i\}$ satisfy the following identities: $\abs{c_0}^2+\abs{c_1}^2 = 1$; 	$\abs{c_0}^2-\abs{c_1}^2 = \braket{\sqrt{\mu}}{-\sqrt{\mu}}$. The states $\{\ket{e_i}\}$ are represented by \cite{lin-lutken}
\begin{equation}
	\begin{split}
		\ket{e_0} &= \frac{1}{\sqrt{\cosh(\mu)}}\sum\limits_{n=0}^{\infty} \frac{\sqrt{\mu}^{2n}}{\sqrt{(2n)!}}\ket{2n},\\
		\ket{e_1} &= \frac{1}{\sqrt{\sinh(\mu)}}\sum\limits_{n=0}^{\infty} \frac{\sqrt{\mu}^{2n+1}}{\sqrt{(2n+1)!}}\ket{2n+1}.
	\end{split}
\end{equation}
This implies that the coefficients $\{c_i\}$ are given by
\begin{equation}
	\begin{split}\label{coeff-ss}
		c_0 &= \mathrm{e}^{-\frac{\mu}{2}}\sqrt{\cosh(\mu)}, \\
		c_1 &= \mathrm{e}^{-\frac{\mu}{2}}\sqrt{\sinh(\mu)}.
	\end{split}
\end{equation}
The coherent states $\ket{\varphi_x,\varphi_y}$ ($x,y \in \{0,1\}$) then have the form
\begin{equation}
	\begin{split}
		\ket{\varphi_0,\varphi_0} &= \ket{\sqrt{\mu},\sqrt{\mu}} = \begin{bmatrix}
			c_0^2\\ c_1^2\\c_0c_1\\c_0c_1	
		\end{bmatrix} \\
		\ket{\varphi_1,\varphi_1} &= \ket{-\sqrt{\mu},-\sqrt{\mu}} = \begin{bmatrix}
			c_0^2\\ c_1^2\\-c_0c_1\\-c_0c_1	
		\end{bmatrix} \\
		\ket{\varphi_0,\varphi_1} &= \ket{\sqrt{\mu},-\sqrt{\mu}} = \begin{bmatrix}
			c_0^2\\ -c_1^2\\-c_0c_1\\c_0c_1	
		\end{bmatrix} \\
		\ket{\varphi_1,\varphi_0} &= \ket{-\sqrt{\mu},\sqrt{\mu}} = \begin{bmatrix}
			c_0^2\\ -c_1^2\\c_0c_1\\-c_0c_1	
		\end{bmatrix}.
	\end{split}
\end{equation}
The representation of $\ket{\varphi_x,\varphi_y}$ is with respect to basis $\{\ket{e_0,e_0},\ket{e_1,e_1},\ket{e_0,e_1},\ket{e_1,e_0}\}$.

\section{POVM operator $F^-_{\text{real}}$}
The POVM operator $F^-_{\text{real}}$ is given by \cite{lin-lutken}
\begin{equation}
	\begin{split}
		F^-_{\text{real}} = (1-p_{\text{dark}})F^-_{\text{mis}}+(1-p_{\text{dark}})p_{\text{dark}}F^?_{\text{mis}}
	\end{split}
\end{equation}
where 
\begin{equation}
	\begin{split}
		F^-_{\text{mis}} &= \begin{bmatrix}
			\frac{a+b+2c+2d+o+p}{8c_0^4}&-\frac{a+b-o-p}{8c_0^2c_1^2}&0&0 \\
			-\frac{a+b-o-p}{8c_0^2c_1^2}&\frac{a+b-2c-2d+o+p}{8c_1^4}&0&0 \\
			0&0&\frac{a-b+o-p}{8c_0^2c_1^2}&-\frac{a-b+2c-2d-o+p}{8c_0^2c_1^2} \\
			0&0&-\frac{a-b-2c+2d-o+p}{8c_0^2c_1^2}&\frac{a-b+o-p}{8c_0^2c_1^2}
		\end{bmatrix},\\
		F^?_{\text{mis}} &= \xi^2\begin{bmatrix}
			\frac{(1+\Omega)^2}{4c_0^4}&0&0&0 \\
			0&\frac{(1-\Omega)^2}{4c_1^4}&0&0 \\
			0&0&\frac{1-\Omega^2}{4c_0^2c_1^2}&0 \\
			0&0&0&\frac{1-\Omega^2}{4c_0^2c_1^2}
		\end{bmatrix}.
	\end{split}
\end{equation}
Note that $\xi=\mathrm{e}^{-\sqrt{\eta}\mu}$, $\Omega=\mathrm{e}^{-2(1-\sqrt{\eta})\mu}$, where $\eta=\eta_{\text{det}}\eta_{\text{lt}}\eta_{\text{atm}}$ (see Tab. \ref{table-parameters}). Take into account that
\begin{equation}\label{elements}
	\begin{split}
		a &= (1-\xi^{1+\sqrt{V}\mathrm{e}^{-\sigma^2_{\phi}}\cos{\delta}})\xi^{1-\sqrt{V}\mathrm{e}^{-\sigma^2_{\phi}}\cos{\delta}},\\
		b &= (\xi^{2(1+\sqrt{V}\mathrm{e}^{-\sigma^2_{\phi}}\cos{\delta})}-\xi^{1+\sqrt{V}\mathrm{e}^{-\sigma^2_{\phi}}\cos{\delta}})\xi^{1-\sqrt{V}\mathrm{e}^{-\sigma^2_{\phi}}\cos{\delta}}\Omega^2,\\
		c &= (\xi^{1+i\sqrt{V}\mathrm{e}^{-\sigma^2_{\phi}}\sin{\delta}}-\xi)\xi\Omega, \\
		d &= (\xi^{1-i\sqrt{V}\mathrm{e}^{-\sigma^2_{\phi}}\sin{\delta}}-\xi)\xi\Omega, \\
		o &= (1-\xi^{1-\sqrt{V}\mathrm{e}^{-\sigma^2_{\phi}}\cos{\delta}})\xi^{1+\sqrt{V}\mathrm{e}^{-\sigma^2_{\phi}}\cos{\delta}}, \\ 
		p &= (\xi^{2(1-\sqrt{V}\mathrm{e}^{-\sigma^2_{\phi}}\cos{\delta})}-\xi^{1-\sqrt{V}\mathrm{e}^{-\sigma^2_{\phi}}\cos{\delta}})\xi^{1+\sqrt{V}\mathrm{e}^{-\sigma^2_{\phi}}\cos{\delta}}\Omega^2
	\end{split}
\end{equation}
The coefficients $c_0$ and $c_1$ are defined in Eq.~\eqref{coeff-ss}. Notice that the \textit{phase turbulence} is involved in $F^-_{\text{real}}$\textemdash exponent $\mathrm{e}^{-\sigma^2_{\phi}}$ (see Appendix \ref{math-single-photon}) takes part in the elements of $F^-_{\text{real}}$. Furthermore, analogous to \eqref{cos-turb}, we have 
\begin{equation}
	\begin{split}
		\langle\sin(\delta+\Delta\phi_{\text{turb}}) \rangle = \mathrm{e}^{-\frac{2\sigma^2_{\phi}}{2}}\sin(\delta) = \mathrm{e}^{-\sigma^2_{\phi}}\sin(\delta)
	\end{split}
\end{equation}
which appears in $c$ and $d$ of \eqref{elements}.

\backmatter

\bmhead{Acknowledgements}
This work is supported by Technical University of Varna under Scientific Project Programme, Grant No. $\text{H}\Pi$6/2025.

\section*{Declarations}
\begin{itemize}
\item \textbf{Funding}.
The article is supported by Technical University of Varna as well as Bulgarian Ministry of Education and Science.
\item \textbf{Conflict of interest/Competing interests (check journal-specific guidelines for which heading to use)}.
The author declares that there is no conflict of interest concerning the publication of this article.
\item \textbf{Ethics approval and consent to participate}.
Not applicable.
\item \textbf{Consent for publication}.
Not applicable.
\item \textbf{Data availability}. 
The data that support the findings of this study are available from the corresponding author upon reasonable request.
\item \textbf{Materials availability}.
The materials that support the findings of this study are available from the corresponding author upon reasonable request.
\item \textbf{Code availability}. 
The code involved in producing results in this study is available from the corresponding author upon reasonable request.
\item \textbf{Author contribution}.
G.B. contributes on his own to all the aspects concerning submitting/publishing the paper.
\end{itemize}


%\bibliographystyle{plain}
\bibliographystyle{unsrt}
\bibliography{sn-bibliography}% common bib file
%\input{sn-article.bbl}

\end{document}